\documentclass[preprint,journal]{vgtc}       % preprint (journal style)

\ifpdf%                                % if we use pdflatex
  \pdfoutput=1\relax                   % create PDFs from pdfLaTeX
  \usepackage{graphicx}                % allow us to embed graphics files
  \DeclareGraphicsExtensions{.pdf,.png,.jpg,.jpeg} % for pdflatex we expect .pdf, .png, or .jpg files
\else%                                 % else we use pure latex
  \usepackage{graphicx}                % allow us to embed graphics files
  \DeclareGraphicsExtensions{.eps}     % for pure latex we expect eps files
\fi%

\graphicspath{{figures/}{pictures/}{images/}{./}} % where to search for the images

\usepackage{microtype}                 % use micro-typography (slightly more compact, better to read)
\PassOptionsToPackage{warn}{textcomp}  % to address font issues with \textrightarrow
\usepackage{textcomp}                  % use better special symbols
\usepackage{mathptmx}                  % use matching math font
\usepackage{times}                     % we use Times as the main font
\usepackage{cite}                      % needed to automatically sort the references
\usepackage{tabu}                      % only used for the table example
\usepackage{booktabs}                  % only used for the table example
\usepackage{xcolor}
\usepackage{amssymb}
\usepackage{subfig}
\usepackage{amsmath}

\preprinttext{To appear in IEEE Transactions on Visualization and Computer Graphics.}

\onlineid{0}

\vgtccategory{Research}
\vgtcpapertype{please specify}

\title{Analysis of the Near-Wall Flow in a Turbine Cascade by \\ Splat Visualization}

\author{Baldwin Nsonga, Gerik Scheuermann, \textit{Member, IEEE}, Stefan Gumhold, \\ Jordi Ventosa-Molina, Denis Koschichow, and Jochen Fr\"{o}hlich}
\authorfooter{
\item B.~Nsonga and G.~Scheuermann are with the Institute of Computer Science, Leipzig University. E-mail:~\{nsonga,scheuermann\}@informatik.uni-leipzig.de.
\item
 S.~Gumhold is with the Institute of Software and Multimedia Technology, TU Dresden.
 E-mail:~stefan.gumhold@tu-dresden.de.
\item
 J.Ventosa-Molina, D. Koschichow and J.~Fr\"{o}hlich are with the Institute of Fluid Mechanics, TU Dresden.
 E-mail: \{jordi.ventosa\_molina, jochen.froehlich\}@tu-dresden.de.
}

\shortauthortitle{Biv \MakeLowercase{\textit{et al.}}:  Analysis of the mean wall flow in a turbine cascade by splat detection}
\abstract{
Turbines are essential components of jet planes and 
  power plants. 
Therefore, their efficiency and service life are of central 
  engineering interest.
In the case of jet planes or thermal power plants, the heating
  of the turbines due to the hot gas flow is critical.
Besides effective cooling, it is a major goal of engineers
  to minimize heat transfer between gas flow and turbine by design.
Since it is known that splat events have a substantial impact on the heat 
  transfer between flow and immersed surfaces, we adapt a
  splat detection and visualization method to a turbine cascade
  simulation in this case study.
Because splat events are small phenomena, we use a direct numerical simulation
  resolving the turbulence in the flow as the base of our analysis.
The outcome shows promising insights into splat formation and
  its relation to vortex structures.
This may lead to better turbine design in the future.
} % end of abstract


\CCScatlist{ % not used in journal version
 \CCScat{K.6.1}{Management of Computing and Information Systems}%
{Project and People Management}{Life Cycle};
 \CCScat{K.7.m}{The Computing Profession}{Miscellaneous}{Ethics}
}

\vgtcinsertpkg

\begin{document}


\firstsection{Introduction}

\maketitle

Turbines are rotating devices converting the energy of a fluid
into a torque and driving an electrical power generator or a compressor in a flight engine,
for example.
The geometry and working principle of a turbine blade are similar to that of an airfoil with the blades radially mounted on an axis and the lift generating a circumferential force driving the rotation of the device \cite{Dick:2015:fund_turbomach}. 
 Often, the gas flow through the turbine is hot because it is downstream of a combustor introducing the energy, like in a flight engine, for example. 
In such a case it is important to maintain a temperature of the blades below a certain limit sufficiently far from the melting temperature of the material constituting the blades, so that heat transfer between the gas flow and the blades has to be minimized.

Since measurements in rotating rows of blades are complicated, a traditional
approach in the field, when desiring to conduct fundamental studies, 
is to consider so-called linear cascades \cite{Todd:1947:WindTunnRes}:
In a wind tunnel with two parallel walls, straight blades are mounted in parallel
from wall to wall.
While far from the side walls the flow is two-dimensional, 
various kinds of secondary flows exist close to the side walls. This is also the case in the full rotating machine, and about 30-50\% of the losses in a turbine are due to these secondary flows \cite{Fottner:1989:turb_blade}.
Studying the flow near the side wall in a linear cascade has, for this reason, been
undertaken in the turbomachinery community, as reviewed in \cite{Coull:2017:Endwall}, for example.
Such studies can be made experimentally in a wind tunnel 
but also numerically by computational fluid dynamics.

Turbulent flows are not entirely irregular. Much research has been devoted to
the identification of ordered building blocks in a turbulent flow, termed 
"coherent structures" in the community. Vortices, near-wall streaks, splats, and antisplats 
are some examples. 
The present paper is concerned with splats and antisplats.
A splat is a region in the flow where the fluid locally impinges on the wall and then spreads 
tangentially along the wall, like an impinging jet, just locally and over a limited duration in time
within the more or less chaotic turbulent flow.
This feature transports fluid from the bulk of the flow towards the wall and can, for example, 
lead to a local and instantaneous increase of the wall temperature
if the wall is at a lower temperature than the bulk flow.
Various other situations exist, where such a feature is of relevance, e.g. with
free surface flows \cite{Perot:1995:splats1}, where gas exchange with 
the atmosphere is of interest.
But also the transport of momentum towards the wall is of relevance in some cases.
If the local flow field is identical to a splat, but with the sign of the velocity vector inverted,
such that the flow leaves the surface, this is named antisplat.
Still, these structures appear in an unsteady way which
can not be determined deterministically on large time scales.

Turning back to the present application from the field of turbomachinery
the following research questions are 
relevant for this type of flow:
(i) Does the flow separate from the blade? Turbulent fluctuations increase the
exchange of momentum between outer flow and near-wall flow and reduce the 
tendency to separate. Splats constitute one of the features contributing to
momentum transport.
(ii) How large is the heat flux between the core flow and the wall?
For the reasons named above, this should be small for a turbine.
Splats do contribute to heat transfer and are thus undesired from this point of view.
While heat transfer was not explicitly computed in the simulation analyzed here,
this can be inferred from the turbulence, which is entirely resolved in the data set.
(iii) How much is the flow near the side walls altered with respect to 
the flow far from the side walls by secondary flows, and what are the 
ordered and the turbulent contributions to this perturbation?

In \cite{Nsonga:2019:splats}, the present authors proposed a method to unambiguously define splats and to extract them in a given flow field.
As it is, this basic method is only applicable to data on Cartesian grids and walls in the Cartesian directions.
Most practically relevant flows, however, feature complex geometries, in particular, with curved surfaces.
Here, the method is now extended to this situation and is applied to a type of flow which is highly relevant due to the economical and ecological impact turbomachines have in modern life.
Motivation and construction of the method are guided by the above research questions.
A quantitative assessment in physical terms is beyond the
scope of this study. But already at this stage, the qualitative assessment
provides important hints for the domain scientists featuring as authors 4 to 6.

%\end{itemize}

\section{Dataset characterization}%[JV]
\label{sec:dataset_characterization}

\subsection{Physical configuration}
The T106A profile is a turbine aft-loaded profile, whose design is aimed at maintaining the suction side boundary layer laminar as long as possible \cite{Ciorciari_Kirik_Niehuis_2014}. 
A brief summary of the main characteristics is as follows. 

The experimental cascade is composed of 7 blades, each with a chord of 100mm separated 79.9mm (pitch measured at the trailing edge). Blades are placed with a stagger angle of 59.28$^{\circ}$ and the blade angle of attack is $\alpha=$37.7$^{\circ}$. With this configuration the axial chord length is $L_{ref}=c_{ax}=$85.88mm. The reference flow speed for the simulation is $U_{ref}=U_{in}=80m/s$. The Reynolds number for this simulation based on these quantities is $Re=5\cdot10^4$.

\begin{figure}[!t]
    \centering
    \includegraphics[trim={20cm 7cm 25cm 8cm},clip,width=0.6\columnwidth]{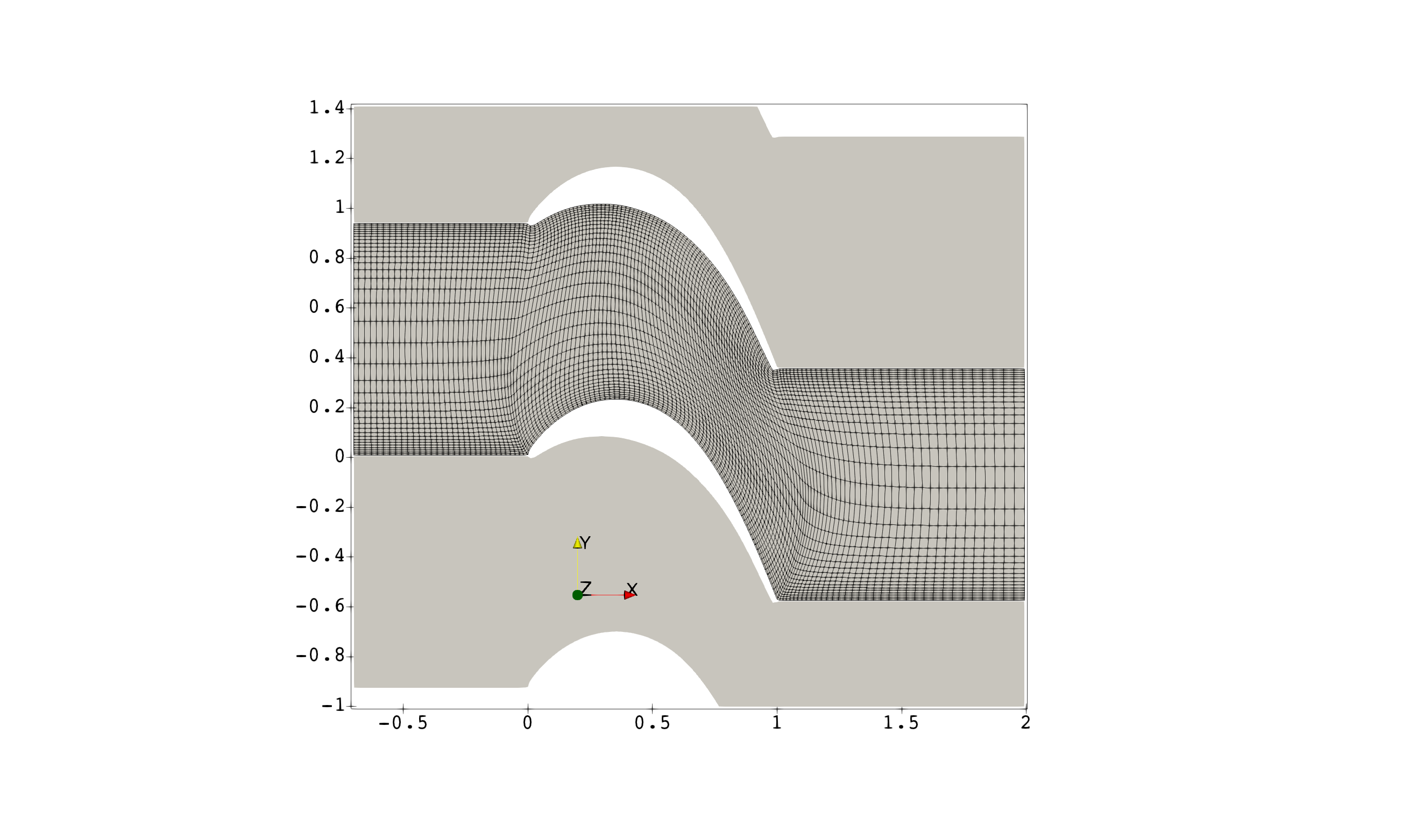}
    \caption{Two-dimensional representation of the physical domain (plane XY). Only one passage between two blades is simulated. The pressure side is at the concave side of the blade (upper curved end of the grid in this figure), the suction side is at the convex side of the blade (lower curved boundary of the grid).  Here only one out of eight grid lines is shown.}
    \label{fig:domain_grid}
\end{figure}

The physical domain simulated corresponds to a subset of the experimental domain, where only the passage between two blades is simulated, as depicted in \autoref{fig:domain_grid}, with periodic boundary conditions in the y-direction. All extensions in space - and, hence, all coordinates - are made non-dimensional using the axial chord length $c_{ax}$.

\subsection{Simulation method for data generation}
The in-house code LESOCC 2 (Large Eddy Simulation on Curvilinear Coordinates version 2) \cite{Hinterberger_Frohlich_Rodi_2008} was used to solve the Navier-Stokes equations in their incompressible non-dimensional form. The code uses a cell-centered finite-volume method with second order central schemes in space and a second order fractional step method time. The solver has been validated in a large number of earlier studies \cite{Wang_Fröhlich_Michelassi_Rodi,WISSINK20081060,Hinterberger_Frohlich_Rodi_2008}.

The data used in the present study is a subset of a study on the endwall flow in low-pressure turbines. The simulation was performed without any subgrid-scale model on a very fine grid resolving all turbulent eddies, thus qualifying as a so-called Direct Numerical Simulation. 
Further analysis on the flow characteristics was presented by Koschichow et al. \cite{Koschichow_Frohlich_Kirik_Niehuis_2014}. 

For the simulation results discussed here, an entirely structured curvilinear hexagonal grid was used with local grid stretching, as shown in \autoref{fig:domain_grid}. The grid consisted of 1154, 290, and 548 cells in the x-,y- and z-directions, respectively. Parallelization was accomplished by domain decomposition using 448 blocks, which for postprocessing were merged into a single block.

\paragraph{Boundary conditions}
A symmetry condition was applied at the profile midspan ($z = C_{ax}$). At the bottom wall ($z=0$) a no-slip condition was imposed. Periodic boundary conditions were set in the pitchwise (y-) direction, upstream ($x<0$) and downstream ($x>c_{ax}$) of the blade surface to account for the cascade consisting of a large number of blades in parallel. This is symbolized by the lower white curved shape which would correspond to the next physical blade in the cascade. This is a standard approach in the field and possible for the flow configurations investigated here. No slip conditions were set at the profile walls ($0<x<c_{ax}$).  A convective outflow condition was set at the outlet plane. 
Regarding the inlet, a turbulent boundary layer was set as inflow. 
The corresponding unsteady inlet condition was generated in a preprocessing step, following the "forcing technique" described by Pierce \cite{pierce_thesis_2001}. The boundary layer thickness was set to $\delta = 0.245c_{ax}$. 
The inflow boundary profile in the domain inlet plane has an angle equal to the profile angle of attack.
Further details are given in \cite{Koschichow_Frohlich_Kirik_Niehuis_2014}.

\begin{figure}[!t]
    \centering
    \includegraphics[trim={5cm 0cm 0cm 8cm},clip,width=0.85\columnwidth]{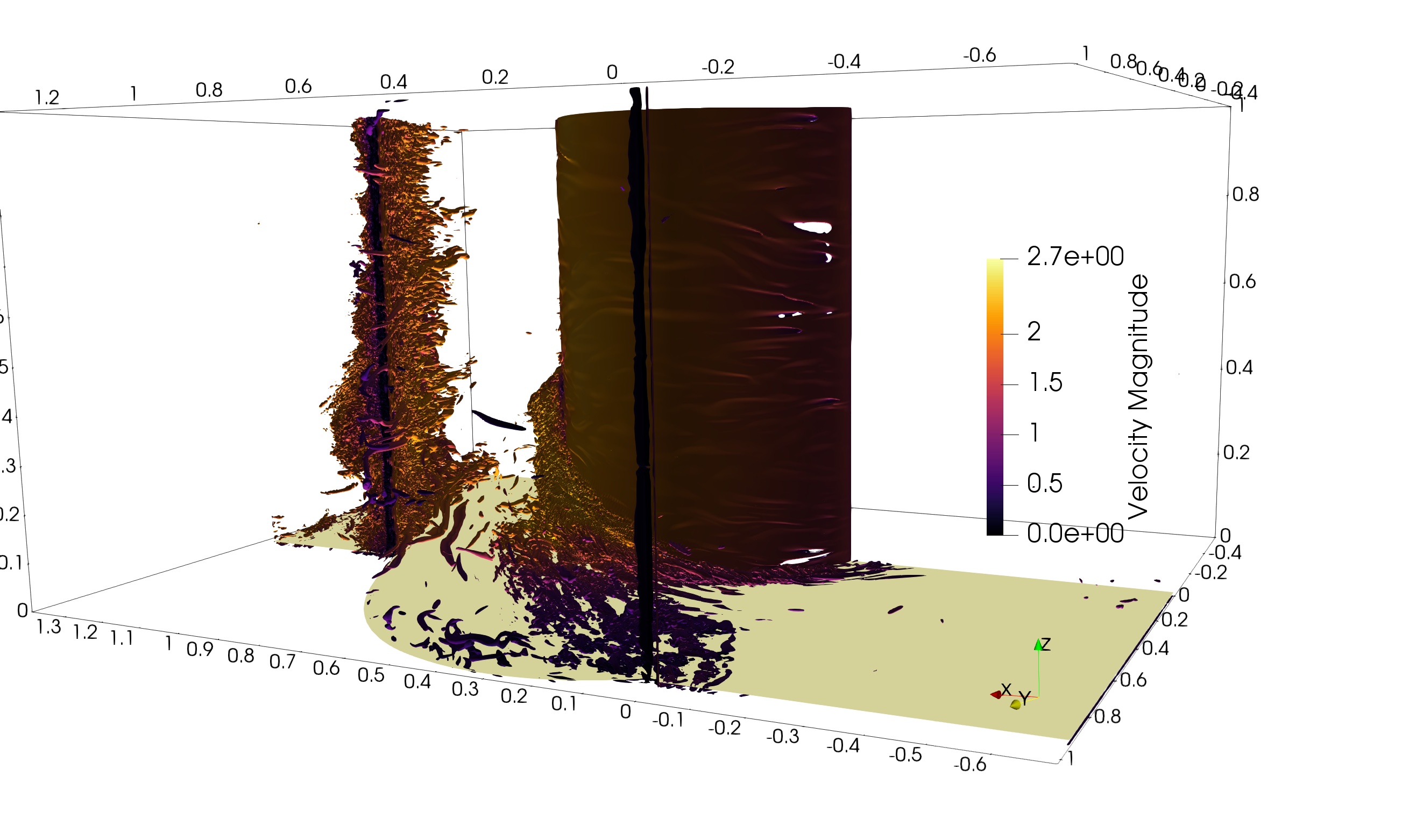}
    \caption{Instantaneous flow field represented through the $\lambda_2=-150$ isocontours. Vortices are colored by the velocity magnitude. }
    \label{fig:inst_lam2}
\end{figure}

\paragraph{Main flow visualization}
\autoref{fig:inst_lam2} shows a perspective view of the vortical structures identified by the $\lambda_2$ criterion. In the figure, secondary flow near the end wall at $z=0$ can be seen, specifically, the flow from the pressure side towards the suction side of the profile.

\begin{figure}[!t]
    \centering
    \includegraphics[trim={4cm 0cm 0cm 8cm},clip,width=0.85\columnwidth]{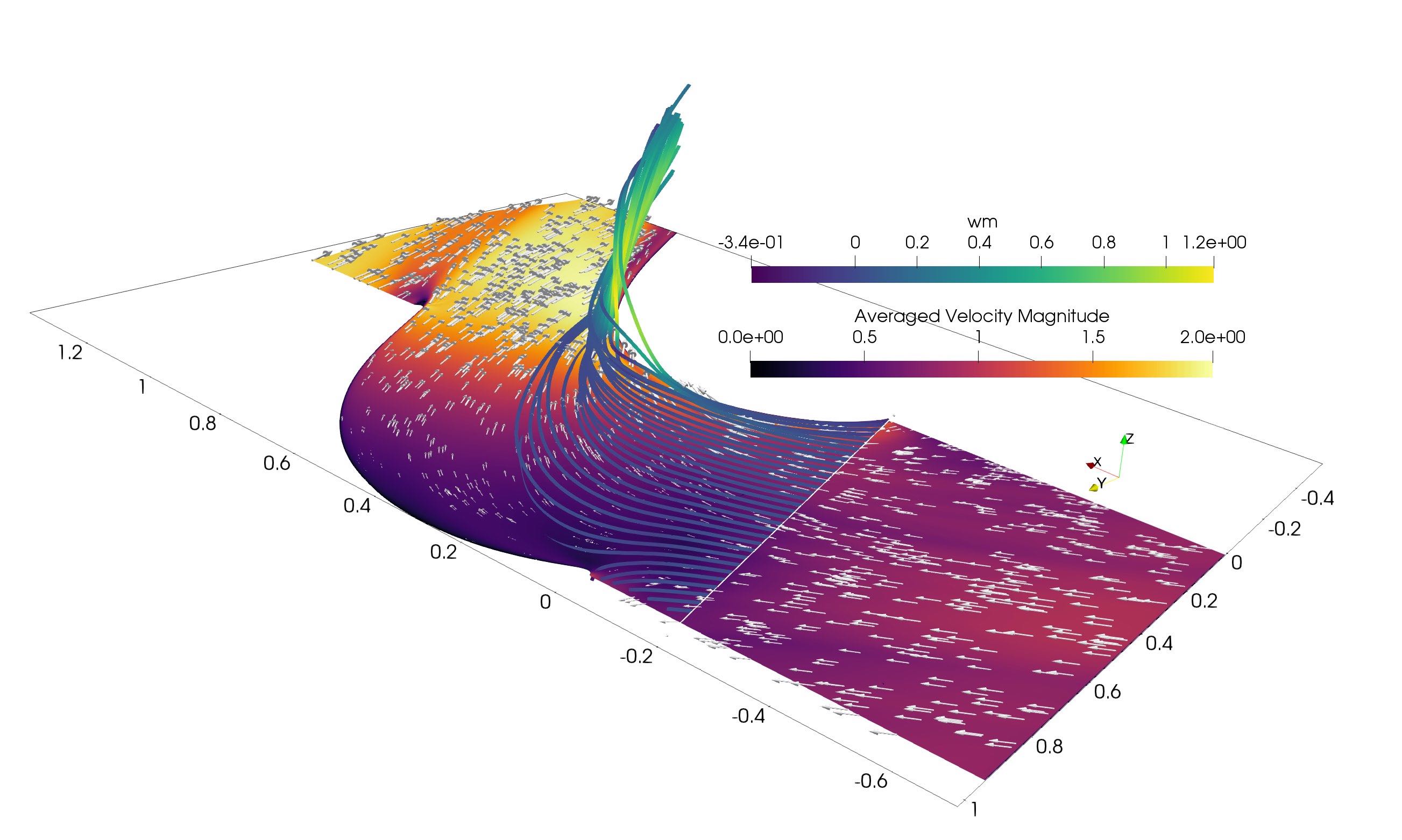}
    \caption{Cut plane at $z=0.02c_{ax}$ showing the time-averaged flow. The color scale indicates the magnitude of the mean flow and superimposed arrows indicate the direction on the plane. Superimposed are 3D streamlines, which start from a straight line at a wall distance of $z=0.02c_{ax}$ in front of the profile and which are colored by the time-averaged vertical (z) velocity, denoted as $wm$ in the legend. 
    }
    \label{fig:average_flow}
\end{figure}

To further visualize the flow field, the time-averaged velocity field near the bottom wall is shown in \autoref{fig:average_flow}.
In the figure, it can be seen that the main flow follows the turbine profile, but there is a marked secondary flow. This can be observed in the figure by the streamlines. These show that the flow is mainly guided by the profile shape, but that there is also a secondary flow path in the vicinity of the bottom wall, where the fluid flows from the pressure side of one profile to the suction side of the neighbor profile.

\section{Related Work}

\subsection{Visualization of turbine flow}

Flow visualization for turbomachinery is commonly achieved through relatively straightforward techniques such as: 
(i) 2D cuts providing velocities, pressure, or other statistical quantities, as color-maps, 
(ii) streamlines, which depict the flow velocities and
(iii) vortex structures by means of the Q \cite{Hunt:1988:q-criterion} or $\lambda_2$ \cite{Jeong:1995:lambda2} criteria.
For an overview of vortex identification techniques, we refer to the survey conducted by G\"{u}nther and Theisel \cite{Guenther:2018:vortex_survey}.

In their case study, Roth and Peikert \cite{Roth:1996:turbomachinery_visualization} discuss various techniques concerned with vortex identification in turbomachinery CFD data.
They stressed that, unlike most other areas of CFD, the flow in these cases is confined to a strongly bent channel which changes its cross-section.
As a result, the flow is mainly governed by geometry rather than vortices. More recent research has pointed out the importance of secondary flow \cite{Coull:2017:Endwall, Giovannini_Rubechini_Marconcini_Simoni_Yepmo_Bertini_2018} and coherent flow structures \cite{Koschichow_Frohlich_Kirik_Niehuis_2014,
Zaki_Wissink_Rodi_Durbin_2010, Zaki_2013} which also is the focus of the present contribution.

\subsection{Background on Splats and Antisplats}

Splats--also referred to as \textit{upwelling}, \textit{upsurge} or \textit{updraught}--- are an important feature of turbulent flows \cite{Pan:1995:splats}.
In his study, Bradshaw \cite{Bradshaw:1981:splats} was the first to theorize these phenomena when studying the pressure Poisson equation in turbulent flows.
Splats and antisplats occur especially in shear-free flows.
Perot and Moin \cite{Perot:1995:splats1,Perot:1995:splats2} defined splats as local regions of stagnation of flow resulting from fluid impinging on a wall.
As fluid cannot penetrate the surface, fluid elements subsequently orient parallel to the wall.
As a result, energy is transferred from the normal to the tangential components.
Perot and Moin extended their studies and found splats in near-wall flows affected by viscous effects.
Additionally, they described antisplats. 
These are reverse phenomena occurring when tangential flow detached and moved in the wall-normal direction.
Perot and Moin argue that the splat-antisplat imbalance results in an energy transference between Reynolds stress components.
Further studies concerned with this imbalance were conducted by Bodart et al. \cite{Bodart:2010:splats}.
In his study, Rashidi \cite{Rashidi:1997:splats}  investigated the relationship between splats, antisplats, and vortices on a free surface.
He found that tangentially rotating vortices coincide with splats and antisplats.
These vortices are associated with the formation of hairpin vortices.
Similar results were observed by Dejoan et al. \cite{Dejoan:2006:splats} in a wall jet.
Also in flow fields with mean shear, an interaction of splats and vortical structures was observed by Hunt~\cite{Hunt:2000:splats}. 

Splats and antisplats play a substantial role in the transport of scalars.
It was shown that they transport heat and pollutants \cite{Tsai:2005:splats, Keating:2004:splats, Wang:2004:splats}.
As heat transfer is particularly crucial for turbines, splats and antisplats are of concern in this field.

The only method for the detection and visualization of splats and antisplats from a velocity field the authors are aware of was introduced by Nsonga et al. \cite{Nsonga:2019:splats}. 
This technique serves as the groundwork for this study.

\section{Splat Detection and Visualization}

\subsection{Splat detection on Cartesian Grids}
\label{sec:splat_detection_cartesian}
In~\cite{Nsonga:2019:splats} splats were defined as regions \(S_R \subset D \times T\), where \( T \subset \mathbb{R}\) is the time domain and \(D \subset \mathbb{R}^3 \) the spatial domain of the flow with boundary \( B = \partial D\), which is assumed to be flat. Splat detection is done on an offset surface \(B^\epsilon\) of the boundary (called offset boundary) at a small positive distance \(\epsilon\) according to the following conditions:

\begin{enumerate}
    \item Non-empty region \(S_R \cap B^\epsilon \neq \emptyset\) is connected
    \item \label{condition_trajectory} \(\forall x \in S_R:  \) Fluid volume moving towards \(B\) along trajectory \(p(\mathbf{x})\): \(p \subset S_R\) is stretched in tangential plane parallel to \(B\) and compressed in normal direction. 
    \item \label{condition_radial} \(\forall t \in T\): it exists a closed curve in region \( S_R \cap B^\epsilon\) with a radial flow pattern.
\end{enumerate}

We now recall the most important details of the method presented in~\cite{Nsonga:2019:splats}, on which we improve in the next section.

\paragraph{Offset Boundary Sampling} 
Assuming a flat boundary composed of points \(B \in \mathbf{x}\), the offset boundary \(B^\epsilon := \mathbf{x} + \epsilon \mathbf{n}\) is defined from the surface normal vector \(\mathbf{n}\) and \(\epsilon \in \mathbb{R}\), which is provided as input parameter to the approach. The offset boundary is represented as a triangular mesh whose vertices serve as seed points for trajectory integration in the next step. A sufficient spatial sampling density is ensured by optional subdivision steps. 

\begin{figure}[!t]
\centering
\subfloat[]{\includegraphics[width=1.2in]{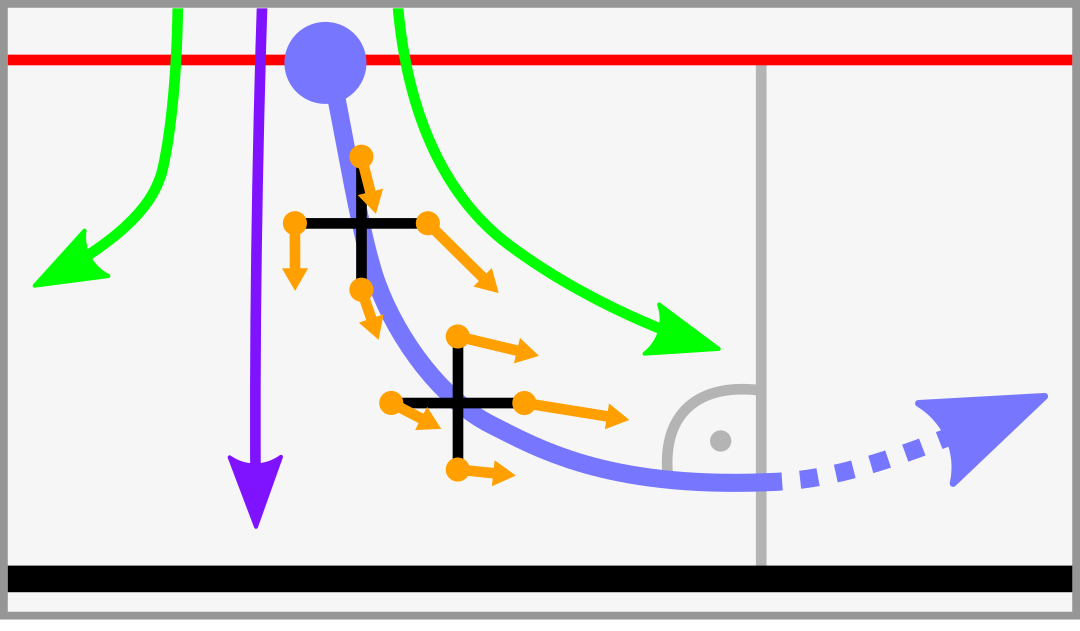}%
\label{fig:lFTLE}}
\hfil
\subfloat[]{\includegraphics[width=1.2in]{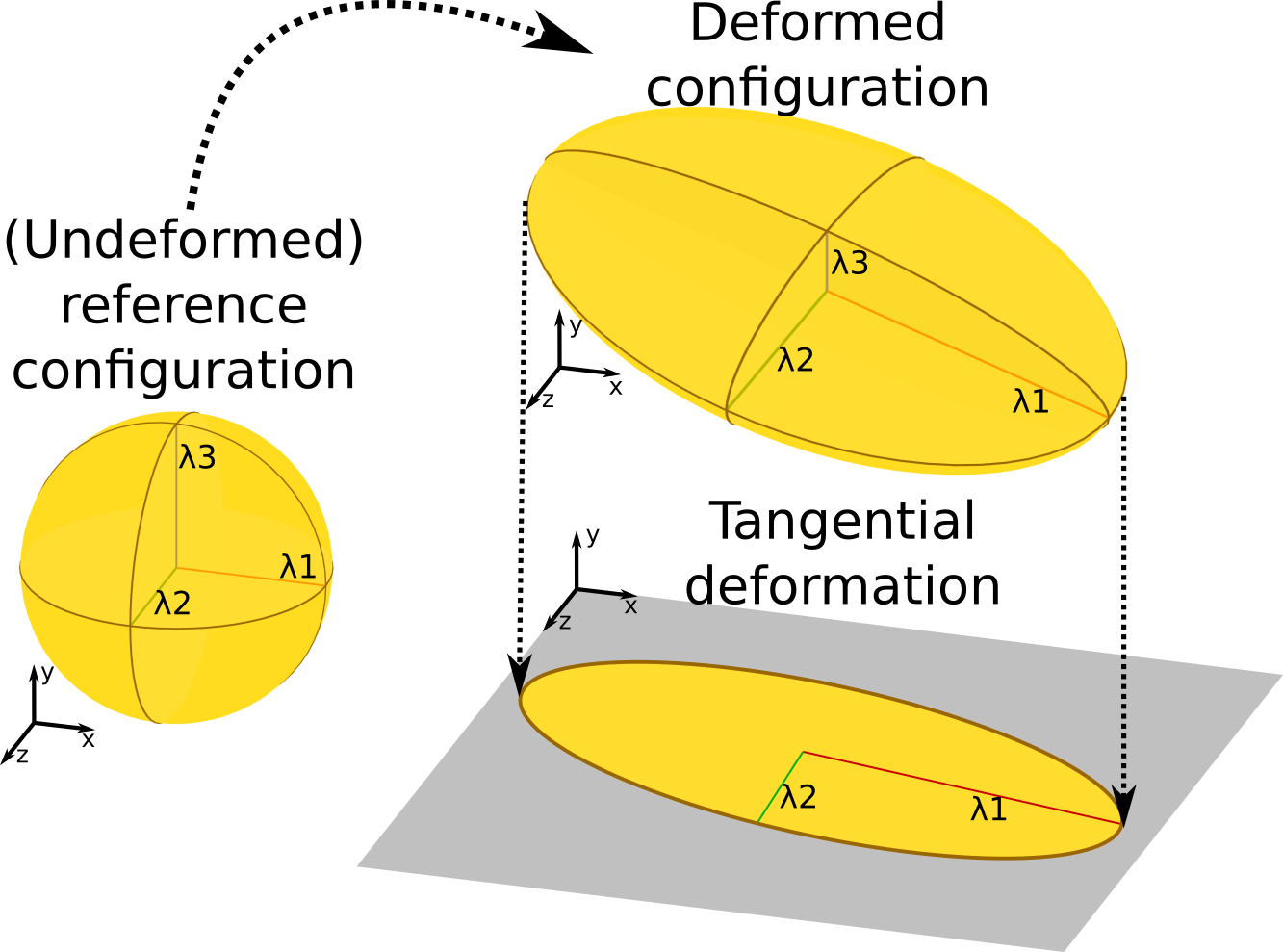}%
\label{fig:projection}}
\caption{(a) Localized surface characterization scheme employed in our method. The blue trajectory shows a path line terminated under the condition stated in Equation \ref{eq:abort}. With the indicated main flow direction (purple) and the attachment flow trajectories (green), the pathline (blue) seeded at the offset surface (red) is sampled for the computation of the spatial gradient. (b) This image shows the tensor glyph of the strain of an isotropic fluid volume in a reference configuration, after advection, and subsequently projected onto the surface. Note that the eigenvalues and eigenvectors vary in every step. Images source: \cite{Nsonga:2019:splats}}
\label{fig:concept}
\end{figure}

\paragraph{Trajectory Integration} 
To find candidate splat regions that fulfill the second condition, trajectories (i.e. pathlines in the instantaneous, frozen flow field) are integrated from the offset boundary vertices.
For splat detection, forward integration is used, and for antisplat detection backward integration.
The temporal distance between integration sample points is provided as input parameter (sample distance) \(\Delta_t \in \mathbb{R}\).
Integration time is limited by the input parameter (maximum integration length) \(\tau \in \mathbb{R}\) and an abort criterion to avoid tracking particles that leave the boundary again (see \autoref{fig:lFTLE}):
\begin{equation}
    \label{eq:abort}
    \forall \mathbf{x} \in p(t) : \mathbf{v}(\mathbf{x}) \cdot \mathbf{n}  \leq 0
\end{equation}
For antisplats the negated velocity is used instead of \(\mathbf{v}\) and the same procedure is applied.
The result of trajectory integration is a list of points 
\(\{s_0, s_1, ... , s_{N-1} \}\) at sample distance \(\Delta_t\) approximating the trajectory \(p(t)\) for splats as well as antisplats.
Note what we employed the Dormand-Prince method \cite{Dormand:1980:RK} for the integration.

\paragraph{Candidate Detection} 
Flow stretching and compression along the integrated trajectories are quantified with the right Cauchy-Green tensor~\cite{Holzapfel:2000:tensor}
\begin{equation}
    C(\mathbf{x}_0) :=  \Psi_{t_0}^t (\mathbf{x}_0)^T \Psi_{t_0}^t (\mathbf{x}_0),
\end{equation}
which measures the square of local change in distance due to deformation
and is computed from the gradient \(\Psi_{t_0}^t (\mathbf{x}_0)\) of a trajectory starting from \(\mathbf{x}_0\) at time
\(t_0\) and integrated until time \(t\). 

The gradient, in turn, is computed according to the technique of localized FTLE~\cite{Kasten:2009:localized}, which, when  performed numerically, is more stable than standard FTLE \cite{Haller:2000:FTLE}. From a sampled
trajectory \(\{s_0, s_1, ... , s_{N-1} \}\), first the gradients of the
velocity field \(\nabla\mathbf{v}(s_i)\) are estimated with central
differences (again using the negated velocity field for antisplats). 
If the computation of the central differences method fails near the boundaries, forward or backward differences method is applied accordingly.
The gradient is finally computed to:
\begin{equation}
\label{eq:psi}
\Psi_{t_0}^t (\mathbf{x}_0) := \prod_{i = N-1 }^0 \mbox{exp}( \Delta_t \nabla\mathbf{v}(s_i) ).
\end{equation}

To be able to distinguish tangential and normal components of the Cauchy-Green tensor, an orthonormal basis 
\(\mathbf{a}\), \(\mathbf{b}\) and \(\mathbf{n}\) is used, consisting of boundary normal \(\mathbf{n}\) and two orthonormal tangent vectors, to define the matrix: \(M := \left[ \mathbf{a} \quad \mathbf{b} \right] \in \mathbb{R}^{3 \times 2}. \)
The Cauchy-Green tensor is then split into the tangential deformation \(C_t(\mathbf{x}_0) := M^T C(\mathbf{x}_0)M \in \mathbb{R}^{2 \times 2}\) and the normal deformation \(C_n(\mathbf{x}_0) := \mathbf{n}^T C(\mathbf{x}_0) \mathbf{n} \in \mathbb{R}\).
Splat candidates are finally defined from the condition
\begin{equation}
    \lambda_{\mbox{min}} \left[  C_t(\mathbf{x}_0) \right] > 1 \wedge C_n < 1,
\end{equation}
where \(\lambda_{\mbox{min}} \left[  C_t(\mathbf{x}_0) \right]\) is the smallest eigenvalue of the tangential deformation tensor (see \autoref{fig:projection}).

\paragraph{Splat Region Detection} 
The candidate detection results in a binary field defined on \(B^\epsilon\). Potential splat regions are extracted from the binary field with an efficient region growing approach that is implemented with a disjoint set data structure and has runtime proportional to the number of seed locations. For details, see~\cite{Nsonga:2019:splats}.

To qualify for a splat, the region should include a sub-region with a radial flow pattern. This is checked by finding all points in the splat region that are critical with respect to the flow field inside the boundary offset surface. If one of these critical points has a positive Poincar\'{e}-Hopf index \cite{Hopf:1927:index} a splat is detected.

As proposed, the method is applied to an unsteady flow field stored over some time interval. Hence, pathlines of fluid elements are computed for splat identification. The method can as well be applied to an instantaneous single snapshot of a flow field, in which case it computes streamlines of this field and can be employed in a similar manner to detect (anti)splats. This involves less data to be treated and is allowed from a practical point of view due to the temporal correlation of the flow field being a solution of the Navier-Stokes equations.

\subsection{Treatment of Curvilinear Grids}
\label{sec:splat_detection_curvilinear}

\begin{figure}[!t]
\centering
\includegraphics[width=0.95\columnwidth]{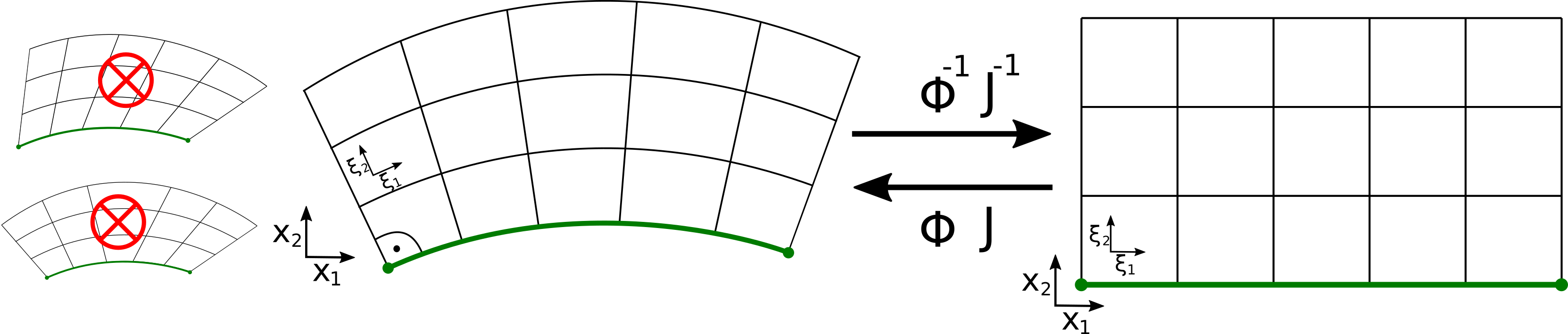}
\caption{ Illustration of the domain transformation applied in this study. It shows the transformations \(J\) and \(\Phi\) between a valid curvilinear grid (physical space) and the uniform grid (computational space). Examples for invalid grids are also shown (left), both containing grid lines with substantial non-orthogonality with respect to the wall.  }
\label{fig:domain_transformation}
\end{figure}

The technique for the detection of splats and antisplats introduced in our prior work \cite{Nsonga:2019:splats} is only applicable along flat surfaces.
The main reason is the computation of the gradient \(\Psi_{t_0}^t (\mathbf{x}_0)\) and the subsequent splitting into tangential and normal components. 
Specifically for the product integral, the spatial gradient is evaluated for each sample point along the trajectory.
On non-flat surfaces, sample points have varying normal vectors.
However, the previously described approach assumes a single normal vector for an unambiguous projection. 
For example, using the normal vector of the seed point evaluates the projected deformation \(C_t(\mathbf{x_0})\) solely on the tangent plane of the seed point.
As the trajectory moves, information of deformation along the surface is lost.
Therefore the original method is not directly applicable in this study.

\paragraph{Grid Requirements}
In the following we generalize the original approach to curvilinear grids.
 Any temporal dependency is not accounted for by the notation in the following, for better readability, but is implied.
For the case of a curvilinear grid the boundary offset surface is defined from the boundary points \(\mathbf{x}_0\) with varying boundary normals \(\mathbf{n}(\mathbf{x_0})\) according to \(\mathbf{y}_0 = \epsilon \mathbf{n}(\mathbf{x_0}) + \mathbf{x}_0\) (see \autoref{fig:domain_transformation}).
In the dataset of our case study, the surfaces of interest correspond to the side wall at \(z = 0\), the suction side of the blade and the pressure side of the blade.
Currently, we do not support grids where the normal direction does not coincide - at least to some degree of accuracy -  with one grid axis. The small inserts on the left of \autoref{fig:domain_transformation} provide two illustrative examples.
Furthermore, grid lines close to the wall need to be at a constant distance from the wall, as highlighted in this figure as well.
Both requirements already correspond to best practice guidelines when establishing the computational grid for the flow simulation, so that they will not incur a true restriction in practice. 

\paragraph{Domain Transformation}
Now the flow field can be transferred into the \textit{computational space}.
The computational space is a uniform grid and commonly used in simulations and various algorithms due to its inherent simplicity. 
For the present case, a grid point spacing of \(\Delta x = \Delta y = \Delta z = 1\) in computational space was chosen.
The domain transformation from computational space to the physical space, curvilinear grid, is denoted as follows: for points as \(\Phi\) and vectors as \(J\).
The transformation for vectors is straightforward:

\begin{equation}
    J = \partial \mathbf{x} / \partial \mathbf{\xi},
\end{equation}
with \(\mathbf{x}\in \mathbb{R}^{3}\) and \(\mathbf{\xi} \in \mathbb{R}^{3}\) being the grid point locations in  physical and in computational space, respectively. 
The transformation of vectors is defined as follows:
\begin{equation}
    \mathbf{v}(\mathbf{x}) = J (\mathbf{\xi}) \mathbf{u}(\mathbf{\xi}), \quad \mathbf{u}(\mathbf{\xi}) = J^{-1}(\mathbf{\xi}) \mathbf{v}(\mathbf{x}),
\end{equation}
where \(\mathbf{u}\) is the velocity vector in computational space and  \(\mathbf{v}\) is the vector in physical space.
In the following, for the sake of simplicity, instead of using \(\mathbf{\xi}\), \(\mathbf{x}\) is used, except for where otherwise noted.
The transformation matrix \(J(\mathbf{x})\) is computed for every grid point using the central differences method.

The transformation of points \(\Phi\) is defined as:
\begin{equation}
    \mathbf{p} = \Phi(\mathbf{q}), \quad \mathbf{q} = \Phi^{-1}(\mathbf{p}),
\end{equation}
 where \(\mathbf{p} \in \mathbb{R}^3\) is a point in physical space and \(\mathbf{q}  \in \mathbb{R}^3\) is the corresponding point in the computational space.
As the seeding and all subsequent steps are performed in computational space, the mapping \(\Phi\)  is only required for the visualization of the streamlines in physical space.
The following approach for \(\Phi\) is applied.
 First, the cell including the point \(\mathbf{q}\) in computational space is located.
 Then, the local coordinates of this point are calculated.
 Subsequently, the local coordinates of the corresponding cell in the physical space are applied to find point \(\mathbf{p}\).
 
 By performing all computations of the splat detection method in the computational space, the boundary surfaces can be treated as flat and the detection technique recalled above can be applied.
Note that due to the transformation, the Jacobian is computed with respect to the adjacent grid points.
As a result, the deformation tensor implicitly takes the curvature of the surface into account.
 
 \paragraph{Summary of method:}
 
The complete approach utilized in this study is as follows:

 Inputs: velocity field \(\mathbf{v}(\mathbf{x})\) defined on a curvilinear grid, maximum integration time \(\tau\), sample distance \(\Delta_t\), offset distance \(\epsilon\).

     (1) Perform the domain transformation for the velocity field: \(\mathbf{u}(\mathbf{x}) = J^{-1}\mathbf{v}(\mathbf{x})\).
     (2) Generate the flat offset boundary surface in computational space and corresponding curved offset boundary in physical space. In this study, slices of the grid are employed as offset surfaces.
     (3) Apply the splat detection method according to \autoref{sec:splat_detection_cartesian} in the computational space.
     (4) Generate streamlines in computational space for the subsequent visualization.
     (5) Map the splat-antisplat scalar field \(\chi(\mathbf{x})\)(see \autoref{eq:chi}) 
     from the computational to the physical space. As we employ slices of the grid in computational and physical space this step is trivial because the indices of the points are equivalent in these spaces or are easily derived. 
     (6) Apply mapping \(\Phi\) to the control points of the streamlines.
     (7) Visualize the resulting splat regions and/or streamlines in the physical space. Details on the visualization scheme are discussed in the following section.

\subsection{Visualization scheme}
\label{sec:visualization_scheme}
\begin{figure}[!t]
\centering
\includegraphics[width=0.7\columnwidth]{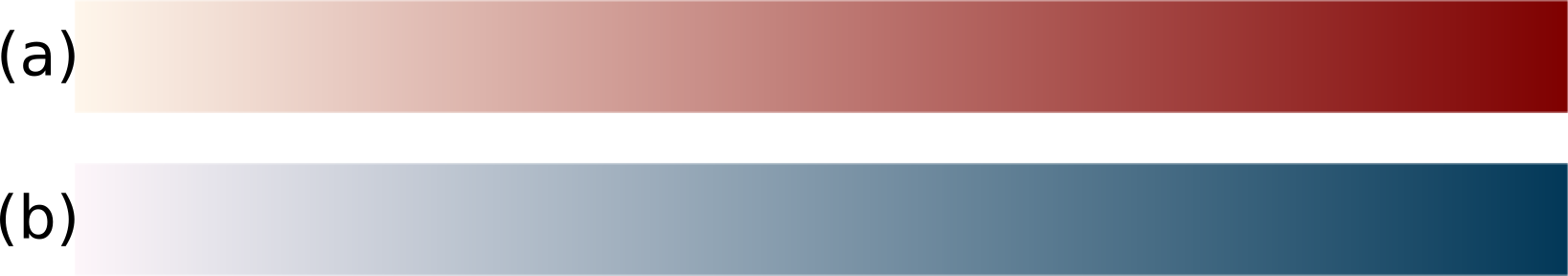}
\caption{ Color mapping scheme for (a) splats and (b) antisplats employed in our structural visualization. The saturation of the coloring increases proportionally to the integration time.}
\label{fig:colormap}
\end{figure}

\paragraph{Surface Visualization}
The surface visualization is performed on the splat-antisplat scalar field \(\chi(\mathbf{x})\) on the offset surface \(B^{\epsilon}\).
Based on the detected splat and antisplat regions \(\chi(\mathbf{x})\) is obtained as follows:
\begin{equation}
   \chi(\mathbf{x}) = 
   \left\{
        \begin{array}{ll}
        +1, & \mathbf{x} \in S_i^{\mbox{splat}} \\
        -1, & \mathbf{x} \in S_i^{\mbox{antisplat}} \\
        0, &  \mathbf{x} \not \in \bigcup S_i  
        \end{array}
    \right .
    \label{eq:chi}
\end{equation}
As stated in \autoref{sec:splat_detection_curvilinear}, scalar fields are mapped to the corresponding offset surface in physical domain.
In the physical domain, the mapping is done directly on the surface (\( \epsilon = 0 \)).
For the visualization a color map \( \chi(\mathbf{x}) \mapsto \{ white, red, blue \}\) as \([-1;0,+1] \to [blue; white; red]\) is employed.
This visualization scheme allows an intuitive assessment of the size and distribution of splat and antisplat events.

\paragraph{Structural Visualization}
Streamline visualization is employed for the depiction of splats and antisplats in the three-dimensional physical space. 
These streamlines are seeded inside the detected splat and antisplat regions of the offset boundary surface \(B^\epsilon\) in computational space.
After that, backward and forward integration is carried out for splats as well as for antisplats.
For the visualization of splat events, the backward integration is the major direction, as it emphasizes where the bulk flow resulting in splat originates.
Forward integration should be limited in this case. 
It should depict the bulk flow impinging on the surface to complete the visualization. 
This step is analogous for antisplats with an emphasis on the forward integration.
Note that the integration times should be chosen according to the case to be studied.
As numerical integration in computational space and the subsequent mapping \(\Phi\) are rather cheap in computation time, this can easily be optimized by the user for a given data set.
The visualization of the streamlines is performed by applying LineAO.
Introduced by Eichelbaum et al. \cite{Eichelbaum:2013:lineAO}, LineAO utilizes ambient occlusion on lines to improve the spatial and structural perception of bulk flows.  
To provide an easy distinction between splats and antisplats as well as a sense of direction, the coloring scheme shown in \autoref{fig:colormap} is employed.

\section{Application to turbine flow}

In this section, the method described in \autoref{sec:splat_detection_cartesian} and \autoref{sec:splat_detection_curvilinear} is applied to the turbine dataset.
The focus is on the near wall flow at the side wall \(\xi_3 = 0\), the pressure side \(\xi_2 = 289 \), and the suction side \(\xi_2 = 0\).
Note that in the computational space of this dataset, these are \(z = \xi_3 = 0\), \(y = \xi_2 = 289\), \(y = \xi_2 = 0\), with the curvature of the blade front being part of the pressure side at \(y = \xi_2 = 289\).
The first step is a decomposition of the velocity field.
On the decomposed data, suitable parameters are derived for the splat detection.
Afterward, an analysis of the sides follows independently.
This analysis is performed on the velocity field \(\mathbf{u}(\mathbf{x})\) in computational space.

\label{sec:application_to_turbine_flowj}

\subsection{Reynolds Decomposition}

As can be recognized from \autoref{fig:inst_lam2} and \autoref{fig:average_flow}, the main flow path \(\mathbf{u}(\mathbf{x})\) is tangential to the turbine blade.
This dominant flow hides small scale features of the flow.
Consequently, splat detection is a challenging task.
To remove the main flow path and reveal the small scale structures, the current approach applies a Reynolds decomposition:
\begin{equation}   \label{eq:reynolds}
    \mathbf{u}(\mathbf{x},t) = \mathbf{\bar{u}(\mathbf{x})} + \mathbf{u'}(\mathbf{x},t), 
\end{equation}
where \(\mathbf{\bar{u}(\mathbf{x})}\) and \(\mathbf{u'}(\mathbf{x},t)\) denote the time averaged flow and the velocity \textit{fluctuations}, respectively.
Adrian et al. \cite{Adrian:2000:analysis_instantaneous_velocity_fields} describe \(\mathbf{\bar{u}}\) as a possible position dependent component with an infinite time-scale. The fluctuation \(\mathbf{u'}\) is shown to reliably reveal small scale flow structures such as vortices in turbulent flow fields.
One weakness, as Adrian et al. \cite{Adrian:2000:analysis_instantaneous_velocity_fields} state, is that the decomposition removes large-scale features associated with the mean flow.
To obtain a comprehensive understanding of the flow, the time-averaged flow \(\mathbf{\bar{u}}\) and the fluctuation \(\mathbf{u'}\) are studied independently.
In the present case this amounts to providing the mean flow field  \(\mathbf{\bar{u}(\mathbf{x})}\), obtained in the simulation by averaging in time at each individual grid point, and a  snapshot of \(\mathbf{u}(\mathbf{x},t)\) at a representative instant in time allowing to compute \(\mathbf{u'}(\mathbf{x},t)\) from (\ref{eq:reynolds}). This flow field is frozen so that the path lines turn into streamlines here without loss of generality.

\subsection{Sensitivity to detection parameters}
\label{sec:parameters}

In this section, the three input parameters for the splat detection method are tested.
This study is performed on the fluctuation field \(\mathbf{u'}(\mathbf{x})\).
The present study is limited to the side wall, as it is sufficient to find good parameter values.

\subsubsection{ Maximum integration time \(\tau\) }
\label{sec:parameters_integration_time}

\begin{figure}[!t]
\centering
\includegraphics[width=0.9\columnwidth]{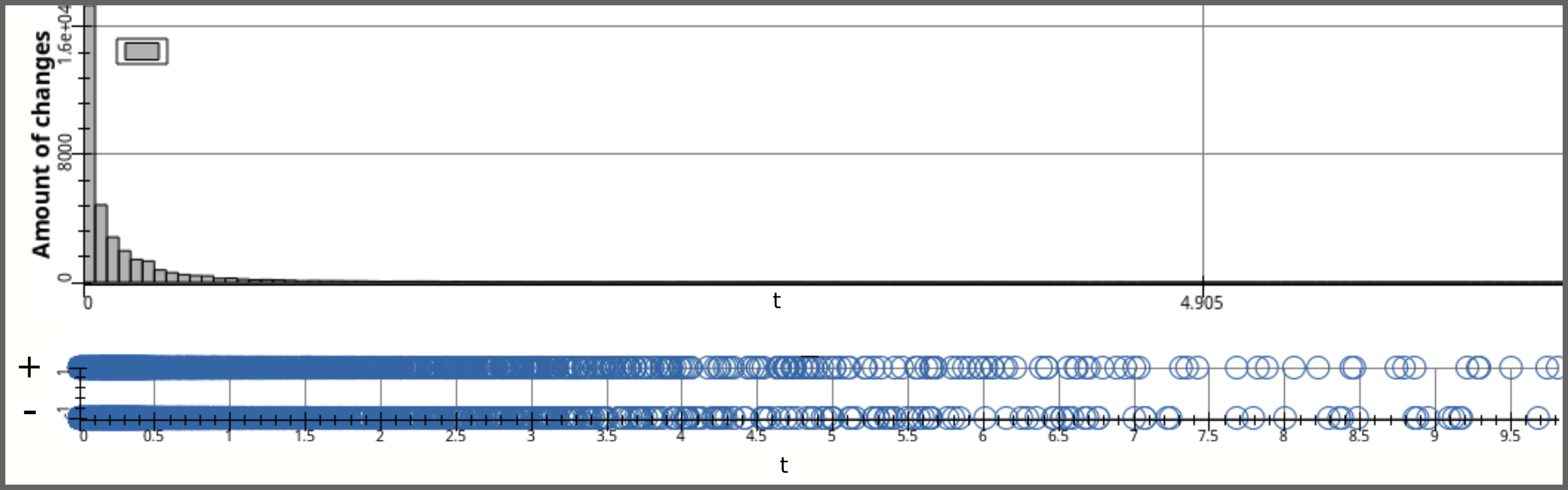}
\caption{ Distribution of slopes of all trajectories. Top: Histogram with the number of slopes occurring at time \(t\). Bottom: Distribution of slopes separated by the changes of a negative to a positive splat candidate evaluation "\(+\)" and the change from a positive to a negative splat candidate evaluation "\(-\)".}
\label{fig:parameters_integration_time}
\end{figure}

In our previous work \cite{Nsonga:2019:splats}, it was stated that the maximum integration time should not be chosen too small, as that would correspond to an instantaneous analysis of the flow.
It should enable trajectories seeded on the offset boundary \(B^\epsilon\) to reach the boundary surface \(B\).
In turbulent flows, high values of \(\tau\) are expected to have minimal impact on the results.
Occasionally trajectories converge towards the boundary.
If not limited, these can lead to artifacts in detected splat regions as shown in our previous study. 

Additionally, the sample length, \(\Delta_t\), and the maximum integration length \(\tau\) have the largest contribution to the computation time.
In the worst case, \(N = \tau/\Delta_t \) samples per trajectory are obtained, each requiring the computation of the spatial gradient and subsequent matrix multiplication.

To find a suitable value for \(\tau\), in our previous work \cite{Nsonga:2019:splats}, a strategy was defined. As a rule of thumb, \(\tau\) is chosen depending on the chosen offset distance.
As prior to the analysis, there is no a priori knowledge over a suitable offset distance, a different approach is employed, described in the following.

For every seed point, the trajectory is integrated until the abort criterion or the previously set maximum integration time is reached.
In this study, an offset of 10 layers and a maximum integration time of \(\tau = 10\) are set.
Based on the obtained results shown later, a higher integration time is found to be unsuitable for this study.
For every sample in a trajectory, a prior evaluation is performed to determine if the splat detection up to this point evaluates a positive splat candidate point or not.
Thus, a binary function is obtained along a trajectory \(e(t)\mapsto \{ true, false \}\).
Subsequently, the behavior of this function is studied in the context of its possible values.
In this process, a differentiation  is made between a \textit{positive slope}, defined as a change from a negative to a positive value, and a \textit{negative slope}, defined as the change from a positive to a negative value.
In turbulent flows, a flow volume gradually performing a deformation can be caught in a flow structure compressing this volume. 
This can lead to a temporary negative splat candidate evaluation according to the definition in \autoref{sec:splat_detection_cartesian}.
As a result, the function can fluctuate.
The goal is to find a time \(t\) which captures all relevant slopes.
The resulting distribution is depicted in \autoref{fig:parameters_integration_time}.
This figure shows that the vast majority of slopes are located in \(t < 1\).
Taking into account all trajectories having at least one slope, \(95\%\) of them have a maximum of 3 slopes over the integration time of the corresponding trajectory. 
After studying the distribution of slopes, we found an integration time of \(\mathbf{\tau = 4}\) to be sufficient.
Only \(0.5 \%\) of the slopes are located above this integration time, which is an acceptable inaccuracy.
This integration time is applied for all following analyses.

\subsubsection{Sample length \(\Delta_t\)}
\label{sec:parameters_sample_distance}

\begin{figure}[!t]
\centering
\includegraphics[width=2.1in]{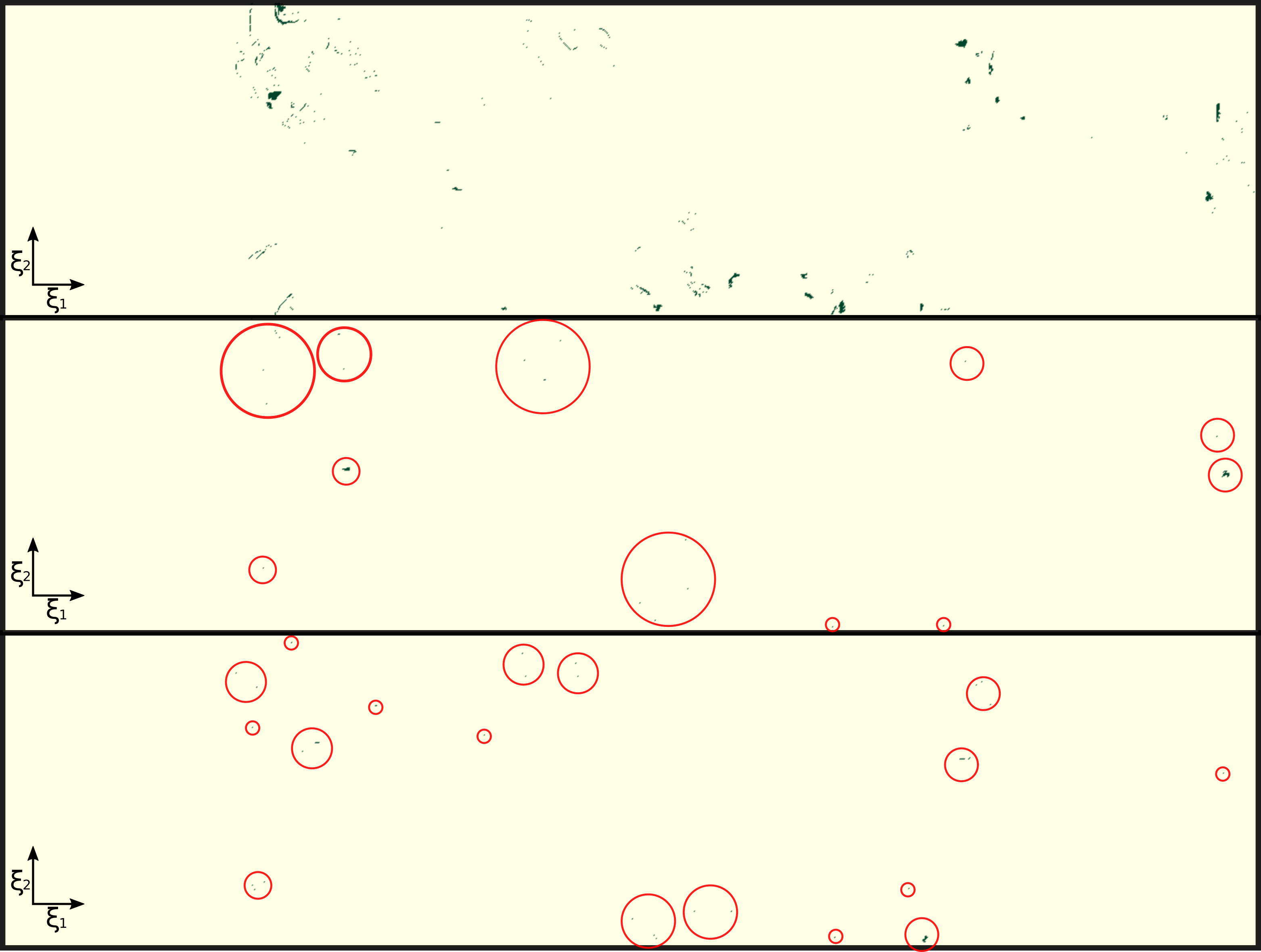}
\caption{ Color map visualizing difference \(d\) defined in \autoref{eq:distance} in the computational space of the (offset) side wall. The coloring scheme is as follows. \(d(\mathbf{x}) = 0: \) yellow. \(d(\mathbf{x}) = 1:\) green. As the sum of distances decreases with increasingly small \(\Delta_t\), we highlight the regions including a point with \(d = 1\) with a red circle. According to \autoref{eq:distance} the following values are shown here. (top): \(\Delta_{t1} = 0.01 \) and  \(\Delta_{t2} = 0.001\). (center): \(\Delta_{t1} = 0.001 \) and \(\Delta_{t2} = 0.0005 \). (bottom): \(\Delta_{t1} = 0.0005 \) and \(\Delta_{t2} = 0.0001 \).}
\label{fig:parameters_sample_length}
\end{figure}

This parameter determines the sampling frequency of the splat detection method.
A low sample frequency can cause artifacts due to aliasing.
Note that \(\Delta_t\) has a significant impact on computation time, see \autoref{sec:parameters_integration_time}.
For this reason, in our previous work \cite{Nsonga:2019:splats}, the following rule of thumb was provided:
\begin{equation}
\label{eq:sample_length}
    \Delta_t = \frac{1}{2} \frac{ \min \{\Delta_x; \Delta_y; \Delta_z\} }{ \max\| \mathbf{u}'(\mathbf{x}, t) \|},
\end{equation}
with the grid point spacing \(\{\Delta_x; \Delta_y; \Delta_z\} \).
As the computations are performed on a uniform grid (computational space) with a spacing of 1, the numerator of \autoref{eq:sample_length} is 1. 
The maximum velocity of the fluctuation is \(\max\| \mathbf{u}'(\mathbf{x}, t) \| \approx 2529\). 
This large value is due to the treatment in computational space with the corresponding scaling by the transformation.
This yields a value of \(\Delta_t \approx 2 \cdot 10^{-4}\) for this criterion.
This is a very low value.
A trajectory reaching maximum integration time \(\tau = 4\) generates \(2 \cdot 10^{4}\) sample points.

As this approach requires significant computational resources, an alternative iterative visual approach is employed here.
It is assumed that starting from a large sample length, the detected splat regions change when reducing the sample length.
On the other hand, the resulting regions should stay constant for sample lengths below the unknown threshold.
Therefore, the splat-antisplat field \(\chi(\mathbf{x})\) is to be computed with different sample lengths \(\Delta_{t1}\) and \(\Delta_{t2}\) with \(\Delta_{t1} \neq \Delta_{t2}\).
We then compute the difference:
\begin{equation}
\label{eq:distance}
d(\mathbf{x}) = |\chi_{\Delta_{t1}}(\mathbf{x}) - \chi_{\Delta_{t2}}(\mathbf{x})|
\end{equation}
Note that antisplats are not included and \(d(\mathbf{x}) \in \{0,1\}\). 
We can then evaluate \(d(\mathbf{x})\) visually.
The goal is to find a sample length with a reasonably small error.
The chosen criteria for the evaluation are (1) very small regions with \(d(\mathbf{x})  = 1\) and (2) these regions should not be clustered.
\autoref{fig:parameters_sample_length} shows three examples for \(d(\mathbf{x}) \) .
Comparing the central to the bottom field in \autoref{fig:parameters_sample_length}  shows only minor differences according to the above criteria.
After this visual evaluation, we conclude that a value of \( \Delta_t  = 0.001 \) is sufficient and will be used in the following applications.

Furthermore, the distribution of velocities was investigated, and it was found that the 99th percentile of velocities is \(\mathbf{u}(\mathbf{x}, t) \approx 300\).
Using this as an input parameter instead of the maximum velocity in \autoref{eq:sample_length} results in a sample length of \(\Delta_t \approx 0.0016\).
This value is of the same order of magnitude as the result of our visual evaluation and supports our choice of the sample length.

\subsubsection{ Offset distance \(\epsilon\) }
\label{sec:parameters_offset}

The offset distance is the most sensitive parameter.
A value too high leads to the method not handling small scale flow features near the wall.
Also, if the offset boundary surface is too far into the flow, vortical structures with a larger distance to the wall may capture the trajectories and trigger the abort criterion.
This could lead to false positive results, especially in turbines as numerous vortical structures can be found.
On the other hand, an offset distance too small may capture effects caused by the viscous effects of the flow on the no-slip boundary.
Also the velocity converges to 0 with \(\mathbf{x} \in B: \|\mathbf{v}(\mathbf{x}) \| = 0\).
The other main point is that the verification of a splat candidate requires the computation of a singular point on the tangential flow field.
In highly turbulent flows, the verification step is susceptible to changes in \(\epsilon\).

In \cite{Nsonga:2019:splats}, we investigated the sensitivity of the procedure with respect to the choice of \(\epsilon\) for an inviscid analytically defined flow.
While the intensity of the (anti)splats was found sensitive to the distance, the detection of the feature itself is not, provided that \(\epsilon\) is below a certain relatively large value. 
In case of a viscous flow, choosing \(\epsilon\) is somewhat more involved since very close to the wall, the magnitude of the velocity vector vanishes due to the no-slip boundary condition.
For turbulent flows, further fine-grain features may appear.

The flow analyzed here features three different walls, the side wall, the pressure side, and the suction side, each identical to a constant index \(\xi_j\), with \(j \in \{1,2,3\}\) of the structured grid employed, so that
the offset boundaries are conveniently chosen to be slices of an index \(\xi_j\) directly mapping to computational space.

To account for the different resolution requirements of this turbulent flow near the three walls, the computational grid of the flow simulation was created with a different step size normal to these walls
with a smaller step size at the side wall compared to the pressure and suction side.
Hence, the same physical distance corresponds to different indices in the grid.
The distance of the offset plane to the wall in computational space, \(\epsilon\), has to be selected based on physical space.
The following procedure was applied here: starting with very small values \(\epsilon\), the parameter
was increased until a value just before the first erroneous features obviously related to vortices appeared.
This yielded a wall distance of \(0.005 c_{ax}\) in physical space, corresponding to
 \(\epsilon = 10\) for the pressure and the suction side and \(\epsilon = 16\) for the side wall.
These values will be applied in the corresponding sections.

\subsection{Splats in the mean flow \(\mathbf{\bar{u}}\)}
\label{sec:application_mean_flow}

\begin{figure}[!t]
\centering
\includegraphics[width=0.8\columnwidth]{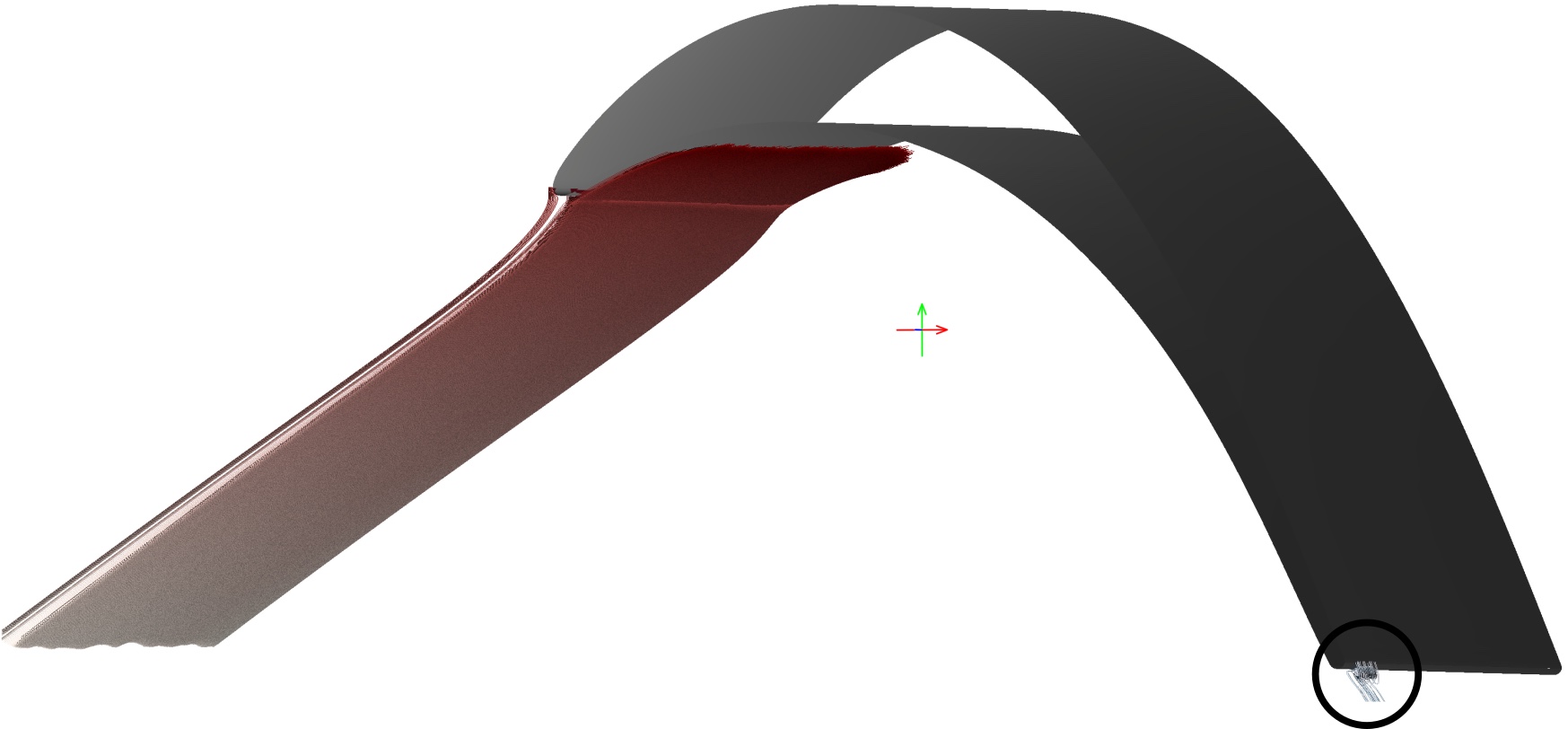}
\caption{Splat and antisplat (circled) visualization performed on the surface of the pressure side in the mean flow \(\mathbf{\bar{u}}(\mathbf{x})\). 
The side wall is located at the rear side of the visualized blade.  }
\label{fig:result_vm_top}
\end{figure}

As a first step, the mean flow \(\mathbf{\bar{u}}\) is analyzed.
The presented method is used together with the previously derived parameters and a two-fold subdivision of the offset surface for improved resolution.
An offset distance \(\epsilon = 10\) at the pressure and suction side and a distance \(\epsilon = 16\) at the side wall are used.

Our method results in a single large-scale splat located at the front of the turbine blade, where the free-stream flow impinges on the turbine blade.
This is illustrated in \autoref{fig:result_vm_top}.
The streamlines, in this case, are integrated backward for a time \(t = 1\) and forward for a time \( t = 0.3\) to showcase origin and further advection of particles involved with the splat.
It is apparent that the flow pattern of the splat is governed by the geometry and the inflow angle.
This large-scale splat is consistent with the expected mean flow behavior.
On the exit side, however, no corresponding large-scale antisplat has been detected.
The antisplat found in the exit of the blade is of small scale and implies a rotation.
This pattern is consistent with the study of Bentaleb et al. \cite{Bentaleb:2012:rounded_step_flow}.
Their study demonstrates that there is a recirculation area behind a rounded step.
As the blade is narrow at the exit, this area is of small scale.
The presence of a vortical structure is verified by \autoref{fig:lambda2}.
In \autoref{fig:result_vm_top} we highlight an antisplat located behind the blade trailing edge.
This antisplat can be related to the low-speed region behind the trailing edge where a small flow recirculation is found.
The comparatively short streamlines generated imply a rather weak flow in that area.
%JF 
Regarding the suction side of the turbine cascade, no splats or antisplats were detected.

\subsection{Splats in \(\mathbf{u'}\) on the suction side}
\label{sec:application_suction_side}

\begin{figure*}[!t]
\centering
\includegraphics[width=1.8\columnwidth]{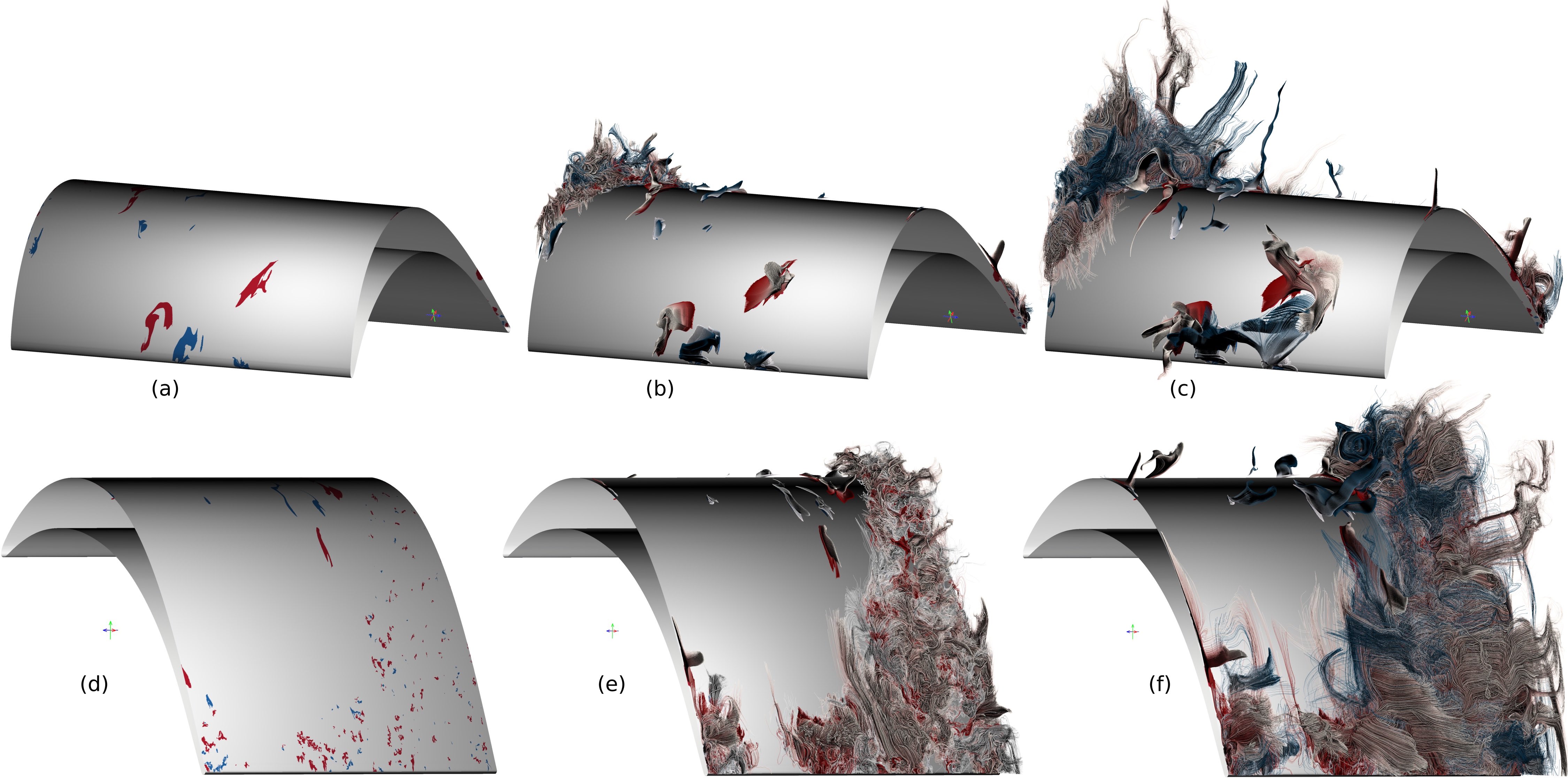}
\caption{ Visualization result of the splat detection approach at the surface of the suction side using the fluctuation field \(\mathbf{u'}(\mathbf{x})\). (a),(b), and (c) are from the inflow perspective and (d), (e), and (f) are from the outflow perspective. (a), (d) 2D visualization of detected splat and antisplat regions as footprints in the offset plane. 3D visualizations (b) and (e) were generated using the forward integration time \(t = 0.3\) and  \(t = 1.0\) for splats and antisplats respectively. The backward integration time is \(t = 1.0\) and  \(t = 0.3\) for splats and antisplats respectively. 3D visualizations (c), (f) were generated analogously with integration times \(t=4.0\) instead of \(t =1.0\). The $x$-direction is indicated by the red arrow of the orientation axes. The green arrow indicates the $y$-direction. The side wall is located at the rear side of the visualized blade.}
\label{fig:results_sunction}
\end{figure*}

In the following, the suction side is analyzed using the fluctuating velocity \(\mathbf{u'}\), with a two-fold subdivision and the parameter values mentioned above.
Results are shown in \autoref{fig:results_sunction}.

The longer integration time was chosen to illustrate the effect of this parameter (\autoref{fig:results_sunction}c,f).
In the present case, the streamlines enter quite far into the domain, revealing unexpectedly large coherent structures.
The shorter integration time provides a clearer picture near the blade with highly turbulent flow in a triangular region along the side wall reflected by multiple irregular (anti)splat events.

Their occurrence and distribution requires a quantitative study which is not in the scope of this work.
The streamlines reveal small and larger vortical flow patterns.
\autoref{fig:results_sunction}f suggests, that the flow resulting in splat events on the suction side near the wall had previously detached from the side wall.

Besides, the increasing lack of coherence with increasing \(x\)-position points towards an increase in turbulence, which is further supported by the large number of small scale splat and antisplat events shown in \autoref{fig:results_sunction}d.
This is a result of the transition of the boundary layer along the blade from a laminar to a turbulent state, as well as the secondary flow. The latter moves fluid from the pressure side to the side wall and from the side wall to the suction side in the form of a large-scale vortical flow pattern, the so-called channel vortex (Fig. \ref{fig:average_flow}).

\subsection{Splats in \(\mathbf{u'}\) on the pressure side}
\label{sec:application_pressure_side}

\begin{figure*}[!t]
\centering
\includegraphics[width=1.75\columnwidth]{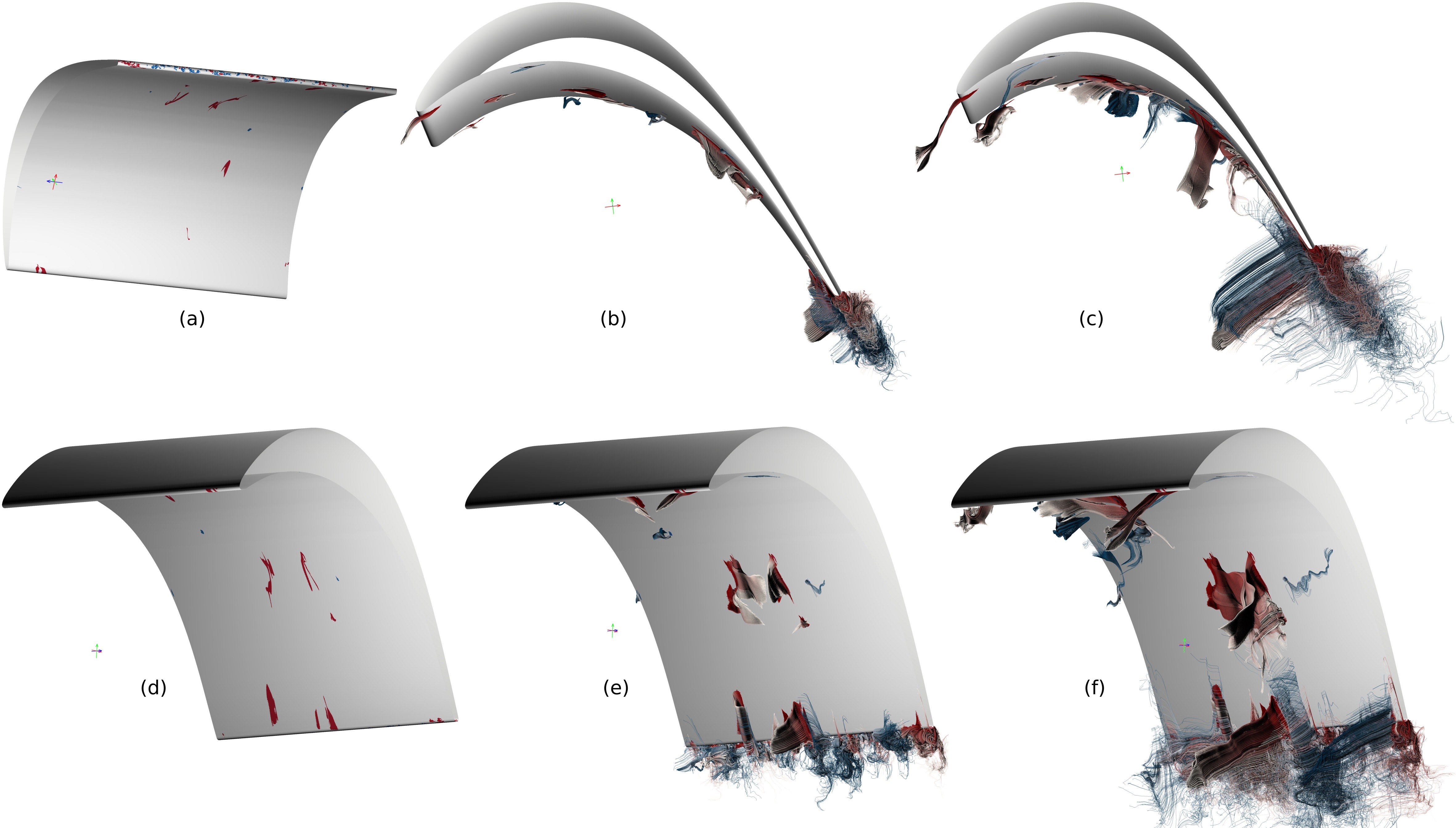}
\caption{ Visualization result of the splat detection approach on the pressure side using the fluctuation field \(\mathbf{u'}(\mathbf{x})\). (a),(b), and (c) are from the inflow perspective and (d), (e), and (f) are from the outflow perspective. (a), (d) 2D visualization of detected splat and antisplat regions. 3D visualizations (b) and (e) are generated using the forward integration time \(t = 0.3\) and  \(t = 1.0\) for splats and antisplats respectively. The backward integration time is \(t = 1.0\) and  \(t = 0.3\) for splats and antisplats respectively. 3D visualizations (c), (f) were generated analogously with integration times \(t=4.0\) instead of \(t =1.0\). Note that the side wall is located at the rear side of the visualized blade. }
\label{fig:results_pressure}
\end{figure*}

Again, this analysis is performed with a two-fold subdivided offset surface, analogous to the previous section with the same parameter values as above.
The visualization results are shown in \autoref{fig:results_pressure}.
The surface area of the pressure side is dominated by splats, indicating a flow transfer from the wall towards the passage. Very few, and also weak, anti-splats events are observed.
The pressure side also shows less turbulence than the suction side.
Additionally, no interaction between splats and antisplats with the side wall is observed.
The splat regions are elongated and suggest flow attachment on the pressure side.
The streamlines resulting in a splat are also subject to a swirling motion when approaching the boundary, suggesting a vortex pattern with core lines parallel to \(\xi_1\).
The streamlines in \autoref{fig:results_pressure}c and f suggest that the origin of these bulk flows is mainly the center of the cascade.
This requires further investigation and is out of the scope of the present work.
The highest density of splats and antisplat can be observed at the exit of the cascade.

\subsection{Splats in \(\mathbf{u'}\) on the side wall}
\label{sec:application_side_wall}

\begin{figure}[!t]
\centering
\subfloat[]{\includegraphics[height=1.75in]{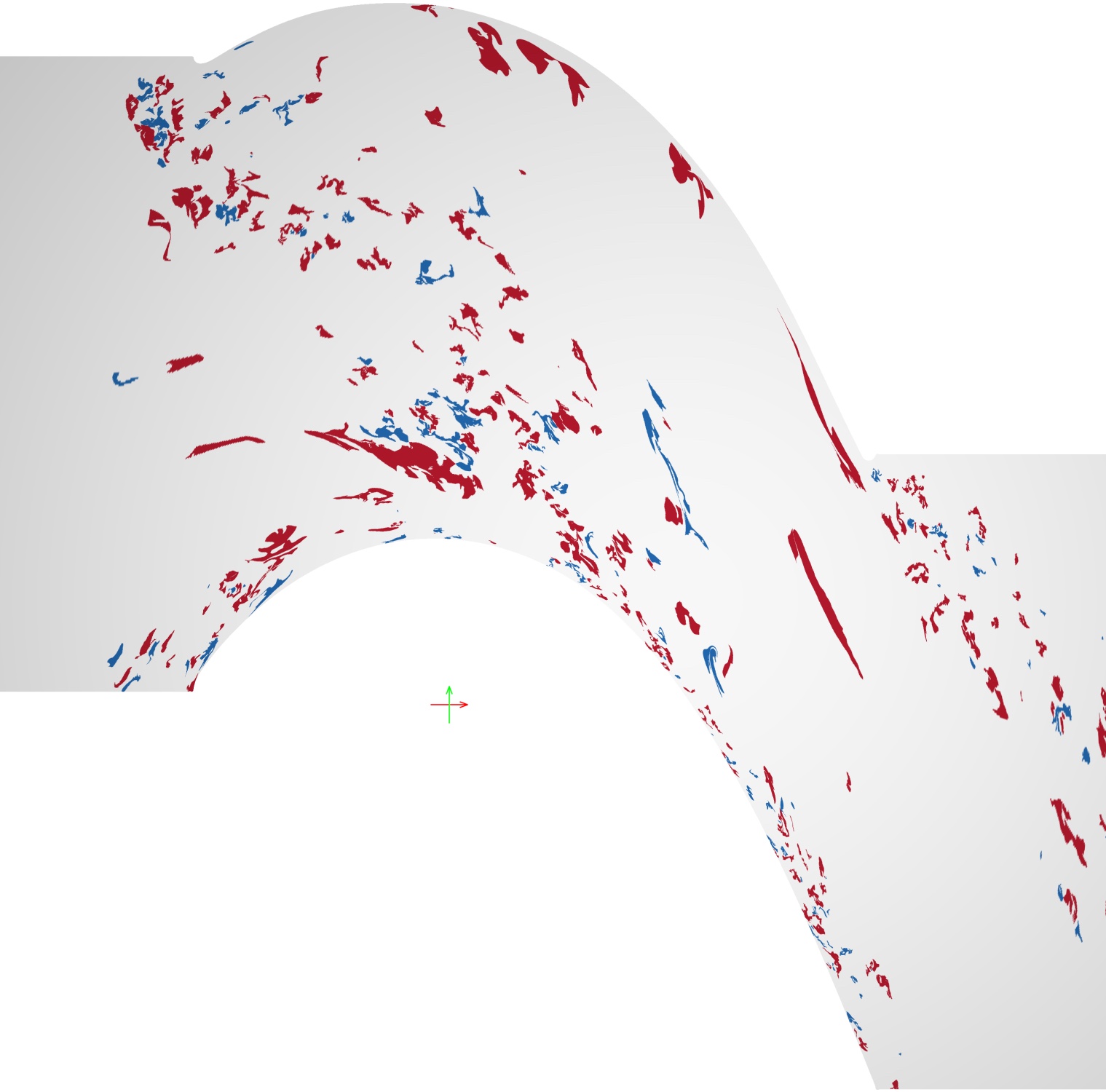}}
\hfil
\subfloat[]{\includegraphics[height=1.75in]{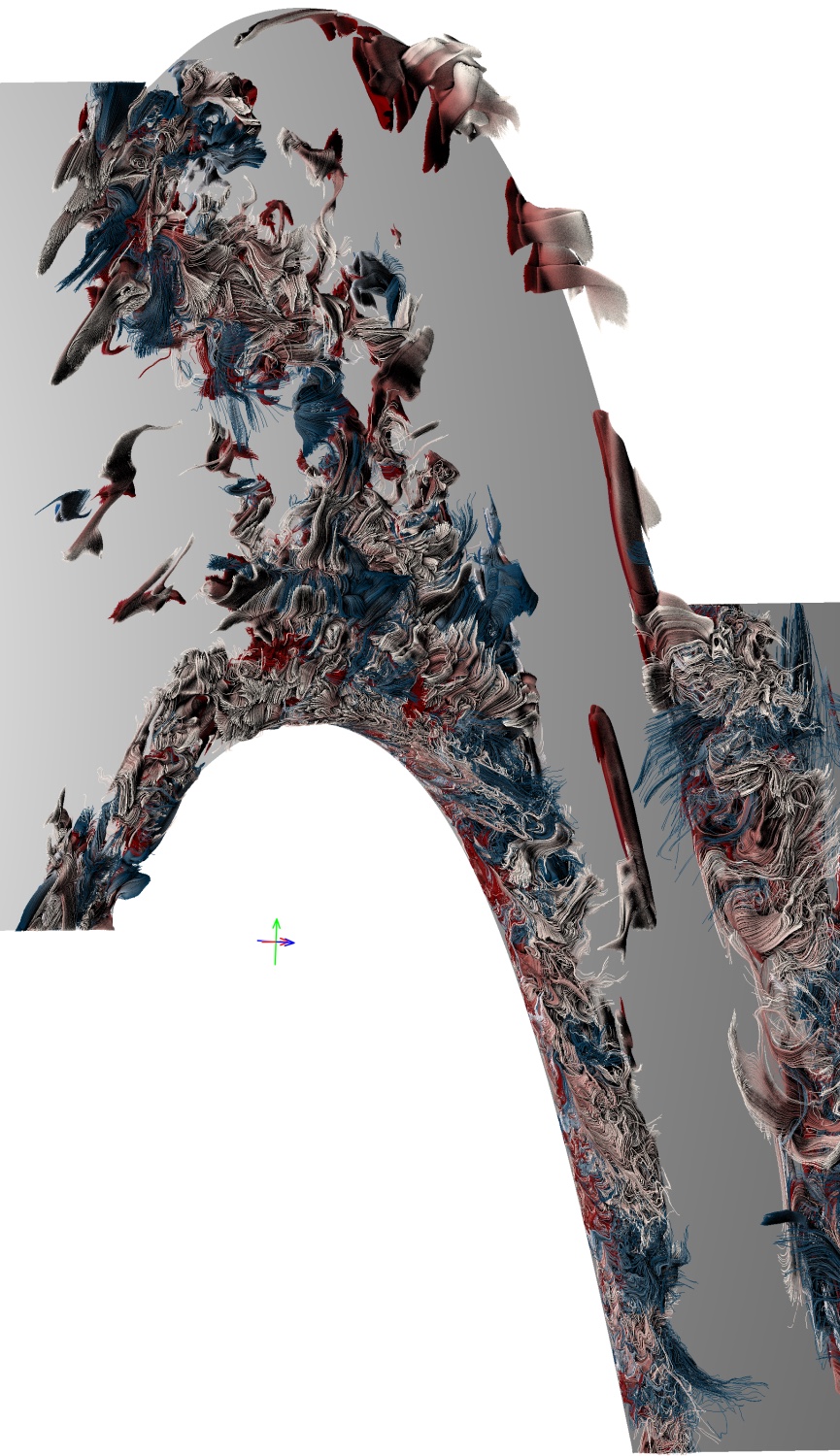}}
\caption{  Visualization result of the splat detection approach at the surface of the side wall using the fluctuation field \(\mathbf{u'}(\mathbf{x})\). 
(a) 2D visualization of detected splat and antisplat regions. 
(b) 3D visualization from the inflow perspective. 
(b) is generated using the forward integration time \(t = 0.3\) and  \(t = 1.0\) for splats and antisplats respectively. The backward integration time is \(t = 1.0\) and  \(t = 0.3\) for splats and antisplats respectively. 
}
\label{fig:results_side_wall}
\end{figure}

This visualization is performed on the two-fold subdivided offset surface with an offset distance \(\epsilon = 16\) in computational space.
The results are shown in \autoref{fig:results_side_wall}.
Our detection approach finds a multitude of splats and antisplats.
The highest densities of splats and antisplats occur near the entrance at the pressure side.
\autoref{fig:results_side_wall}a shows that the majority of splats and antisplats near the suction side are small in scale.
Also, a large quantity of splats is linearly distributed between the front of the pressure side and the exit of the suction side.
This distribution is virtually perpendicular to the entrance angle of the flow.
We assume that a secondary flow in \(y=-1\) direction is responsible for this distribution.
Along the corner between the pressure side of the blade and the side wall, splat events emerge. 
The main focus in a further investigation of the flow should be on the interaction of the side wall with the suction side and the splats/antisplats distributed on the line between pressure side entrance and suction side exit.
These splats may contain information about heat transport.

\section{Discussion}

\begin{figure}[!t]
\centering
\subfloat (a){\includegraphics[height=1.23in]{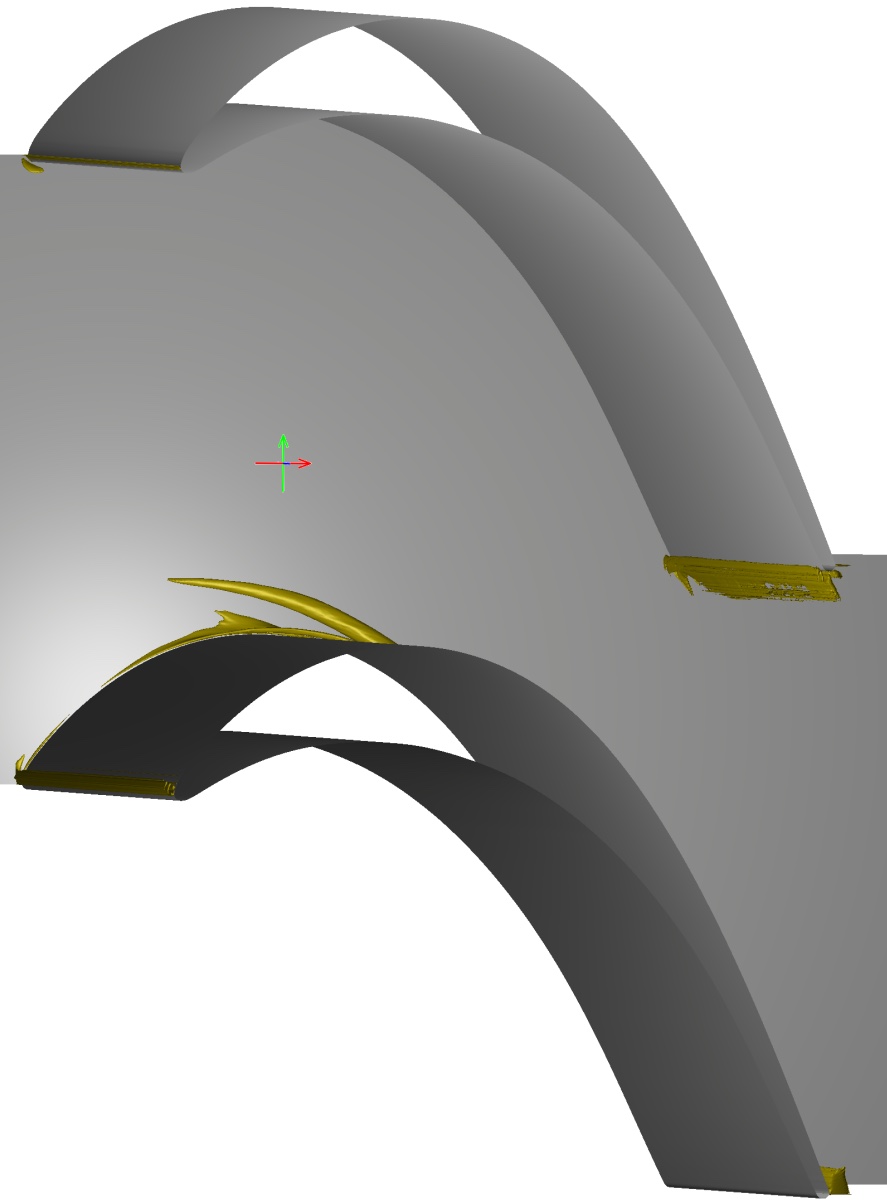}}%
\hfill
%\subfloat (b){\includegraphics[width=0.27\columnwidth]{pictures/vm_lambda_2.jpg}}%
%\hfill
\subfloat (b){\includegraphics[height=1.23in]{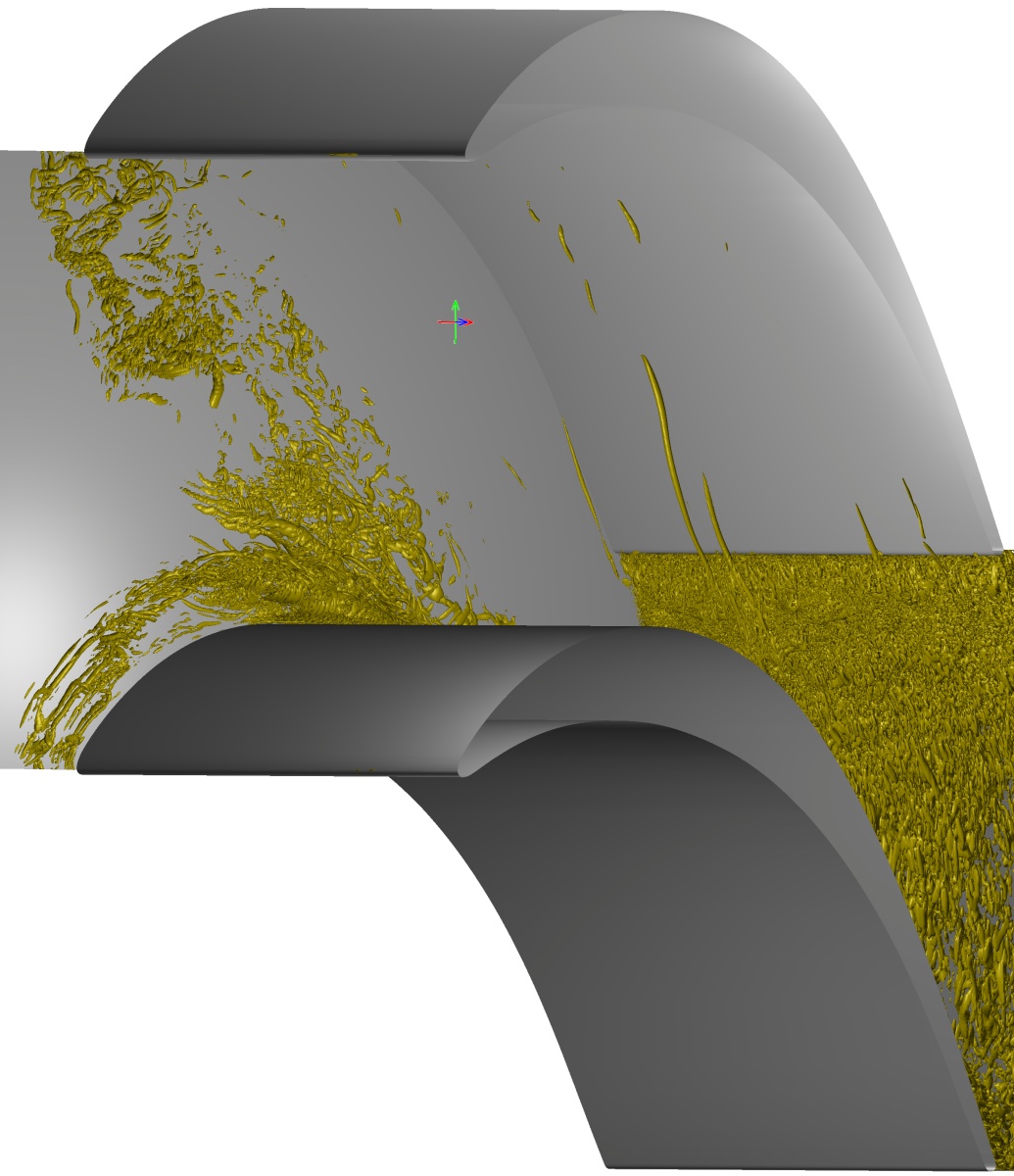}}%
%\hfill
\subfloat (c){\includegraphics[height=1.23in]{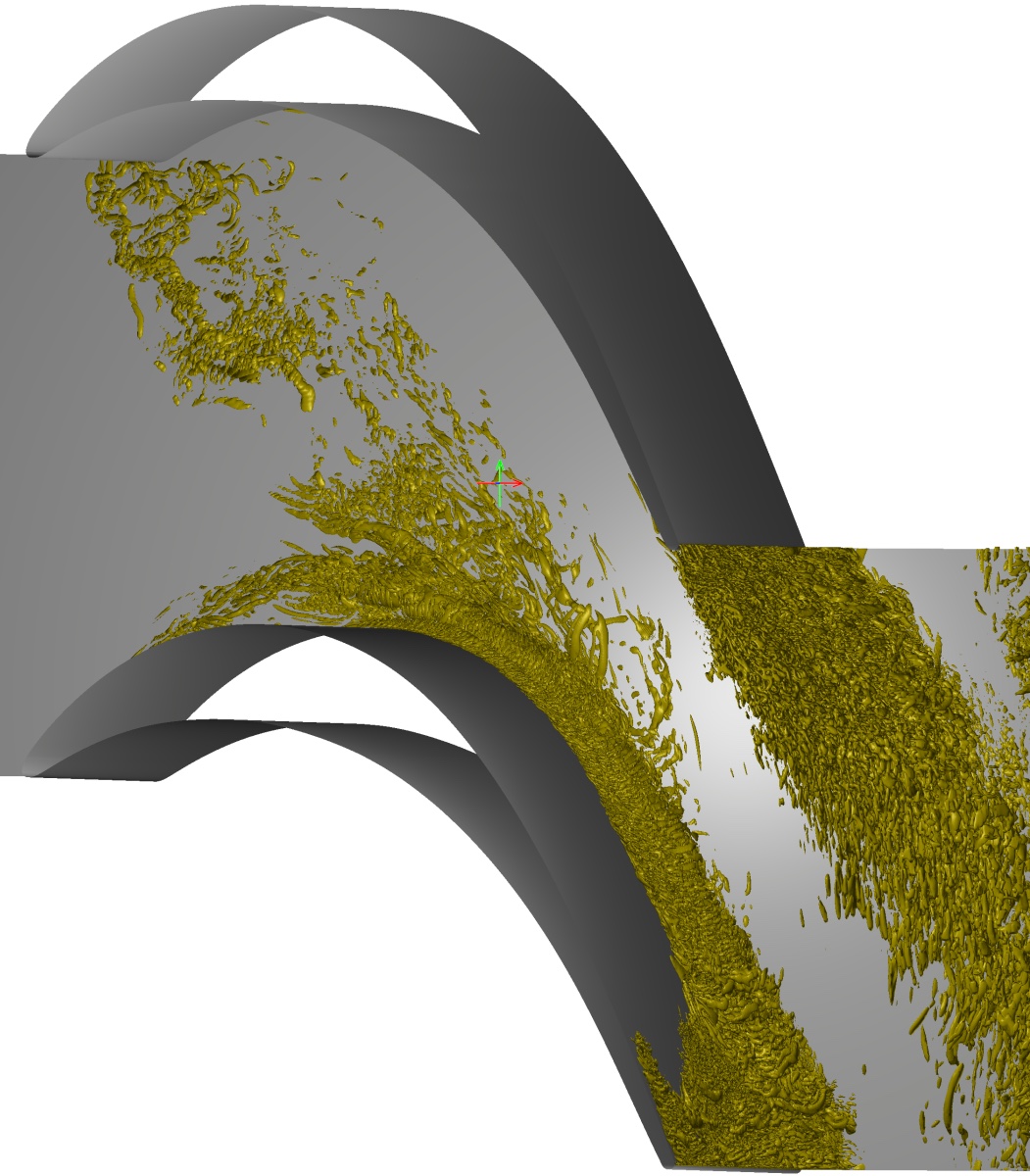}}%
\caption{ Visualization of \(\lambda_2 = -2000\) isocontours (yellow). (a) Evaluation for the mean flow  \(\mathbf{\bar{u}}(\mathbf{x})\). (b) Evaluation based on the velocity fluctuations \(\mathbf{u'}(\mathbf{x})\), (c) the same isosurface as (b), viewed from a different angle.} 
\label{fig:lambda2}
\end{figure}

\subsection{Discussion of findings}
In this section we discuss the results of \autoref{sec:application_mean_flow},\ref{sec:application_suction_side}, \ref{sec:application_pressure_side}, and \ref{sec:application_side_wall} with respect to the research questions stated in the introduction.

\paragraph{(i)} \textit{Does the flow separate from the blade?}

Separation is associated with antisplat events - in the mean as well as in the fluctuations.
Our analysis shows that antisplats occur in the fluctuations, on small scales in an interplay with splats.
They are the result of vortical flows alternating between splatting and separating flows from the blades due to turbulent fluctuations.
A sustained separation of the mean flow from the blade is not observed here.

\paragraph{(ii)} \textit{How large is the heat flux between the core flow and the wall?}

As already stated, the heat transfer has not been simulated in this dataset.
Yet, splats are associated with local heat transfer in the turbulent regime, where
convection is the main cause of heat transfer, rather than diffusion.
The streamlines associated with the splat and antisplat regions are a suitable indicator of whether heat is transferred from the core flow to the side walls.
Examining \autoref{fig:result_vm_top}, \ref{fig:results_sunction}, \ref{fig:results_pressure}, and \ref{fig:results_side_wall} results in several regions of interest.
The first region at the front side of the blade is the stagnation point region (\autoref{fig:result_vm_top}) which is a feature of the mean flow and well understood.
Another region where further studies should focus on is the suction side of the blade. 
Our method shows that an extensive interaction between the core flow and the near wall flow exists in that region (\autoref{fig:results_pressure}c and f), which is highly influenced by the side wall.
The pressure side also features such regions (\autoref{fig:results_pressure}c and f), but with barely any impact of the side wall.
This is important information for the application since regions of the wall with increased numbers of splats will be subject to an increased heat transfer. This may require additional cooling of the blade, internally or by so-called film cooling, where a colder fluid is ejected through holes in the surface shielding the blade against splats of hot fluid.
 On the side wall, a pronounced region of splats is detected along the horseshoe vortex starting at the leading edge (upper end in  \autoref{fig:results_side_wall}) and extending to the apex of the suction side. If film cooling is installed at the side wall, this will be the region where to put the ejections.

\paragraph{(iii)} \textit{How much is the flow near the side walls altered with respect to the flow far from the side walls by secondary flows, and what are the ordered and the turbulent contributions to this perturbation?}

We can perceive the impact of secondary flows on the surface of the suction side and the side wall in the fluctuation flow field, see \autoref{fig:results_sunction} and \ref{fig:results_side_wall}.
A significant amount of small scale splats and antisplats can be found on these sides, related to the secondary flow in the cascade.
In an application, these yield larger heat fluxes from hot fluid towards the cold wall. 
The distribution of the resulting splats and antisplats on the suction side is consistent with \autoref{fig:average_flow}.
As we found no additional splats and antisplats in the mean flow at the side walls with respect to a secondary flows, this contribution is small.

\subsection{Discussion of method}

\subsubsection{Variants of the method}

With the results presented in the previous section, splat detection in conjunction with a domain transformation is shown to work well on curvilinear grids and curved offset surfaces, even in a complex flow as the present one.
The method generates two types of results, 2D surface plots providing the footprint of (anti)splats and 3D pathlines or streamlines. Both have their respective merits in such a complex flow. The 2D plots are easy to analyze and provide a condensed perspective on the occurrence of these features. The 3D streamlines relate the (anti)splats to the interior of the flow field and connect them to vortical structures. In this way, the origin and the reasons for their occurrence can be addressed and analyzed from a fluid mechanical perspective.

\subsubsection{Comparison to regular approaches}

A standard visualization tool for complex turbulent flows is the $\lambda_2$ method \cite{Jeong:1995:lambda2}, the most prominent member of an entire family of vortex extraction criteria, also containing the $Q-$criterion, the discriminant method, etc. All these determine a scalar and identify the presence of a vortex with values of the scalar beyond a user-specified threshold so that an isosurface is employed to extract the respective regions. This was done in \autoref{fig:inst_lam2} by the application experts using their standard procedure and is repeated in \autoref{fig:lambda2} with a style similar to the splat analysis performed in \autoref{sec:application_to_turbine_flowj}.

A comparison of these graphs with \autoref{fig:results_sunction}-\ref{fig:results_side_wall} shows that splat detection is not just another tool for identifying certain vortices but a genuinely different approach, providing complementary information.
This is highlighted by the results for the mean flow (\autoref{fig:lambda2}a vs. \autoref{fig:result_vm_top}) and the fluctuations (\autoref{fig:lambda2}b, c vs. \autoref{fig:results_sunction}, \ref{fig:results_pressure}).
For example, the $\lambda_2$-isosurfaces are close to but remote from the side-wall boundary in \autoref{fig:lambda2}b along the front part of the suction side with no vortex structures identified directly at the blade surface near the leading edge. 
In contrast, \autoref{fig:results_sunction} shows splats in $\mathbf{u'}$ near the leading edge of the blade resulting from impinging free stream turbulence. 
Another example is the pressure side with two splats in the middle (\autoref{fig:results_pressure}d, e) and two small vortices along the blade surface (\autoref{fig:lambda2}b).

Some differences of the present method with respect to the cited vortex criteria should be highlighted here.
\textit{(1)} There is no consensus over how a vortex is defined in the fluid mechanics community. By visualizing streamlines utilizing LineAO, we provide an easy and intuitive way of visually identifying vortex structures related to (anti)splats. 
\textit{(2)} Vortex detection, e.g., via \(Q\) or \(\lambda_2\) criteria requires the choice of a detection level which
is delicate and may not be uniform over an entire flow.
The presented method allows a direct derivation of suitable parameters, as shown in \autoref{sec:parameters}. The choice of the offset distance \(\epsilon\) requires user assessment but is uncritical. 
An improper value for \(\epsilon\) can easily be detected by examining the (anti)splats in the 3D visualization.
The integration time for the streamlines or pathlines also is a user-defined parameter (\autoref{fig:results_sunction}-\ref{fig:results_side_wall}). Changing this value does not change the essential part of the feature, as with the \(\lambda_2\) criterion, but just increases the length of the detected feature.
\textit{(3)} Splats and antisplats can be generated other than from vortices. They are still detected by our method.

Commonly used streamline techniques require extensive knowledge of the flow, as the seeding for the streamlines must be proficiently made to obtain relevant results. The drawback of this process is that events in unexpected regions can be easily missed. The present method is exempt of this issue as, here, streamline seeding is a result of the detected regions and does not require further user input except for the length of the visualized streamlines.

Overall, the proposed method offers a systematic and, most of all, an automated approach to visualize splats and antisplats in complex laminar and turbulent flows. This was highlighted here by the mean flow and the turbulent fluctuations. Otherwise, such an extraction would require extensive manual work of the CFD user to create a multitude of streamline visualizations and search for zones where possible splat or antisplat events could be happening. Then, together with prior knowledge of the user about 
the flow in question, candidate regions could be identified to look for splats/anti-splats structures and dedicated streamline representations would be created as well as local maps to provide an impression of the impingement or separation. 
The present work shows that the developed tool constitutes a substantial enrichment of the arsenal of methods available
to the CFD researcher suitable to address issues of flow impingement so far rarely being discussed in detail due to the
lack of a suitable and user-friendly tool.

\subsubsection{Limitations of the method}
The major drawback of this approach is the 3D visualization scheme in regions of strong turbulence.
Small scaled splat and antisplat interactions result in a high amount of clutter.
Therefore, the interpretation of the results becomes very challenging in cases with pronounced turbulence.

Furthermore, the method does not provide consistent handling for corners.
Still, analyzing the three independent smooth surface areas together circumvents this issue.
Besides, the error related to the computation of the transformation matrix \(J\) adds an inaccuracy which is amplified in non-smooth regions of the domain, e.g., corners.
Nevertheless, this error is insignificant in the present study, as we exclusively visualize splats/antisplats of the (smooth) blade and the side wall.

\subsubsection{Performance}

The computation times where obtained on a Linux system with Ubuntu 18.04 LTS, 32GB of RAM, and 32x Intel(R) Xeon(R) CPU E5-2630 v3 @ 2.40GHz (hyperthreaded).
The domain transformation is straightforward with a time complexity of \(O(n)\).
After a two-fold subdivision the number of seed point is approximately \(n = 5.3 \cdot 10^6\) and \(n = 10.1 \cdot 10^6\) for the side wall surface and the pressure/suction side surfaces, respectively.
Note that the splat detection algorithm was applied to the complete side wall, including the periodical sections.
The worst case time complexity for the computation and evaluation of the deformation tensors is \(O(n \tau / \Delta_t )\).
Note that the performance of the numerical integration and the abort criterion of the integration highly depend on the flow.
The splat and antisplat detection for the suction wall was performed in approximately 27min.
For the pressure side and the side wall, we measured approximately 32min and 26min, respectively. 
Mean flow computation took approximately 28min per side.
For this case study, we think the computation time to be reasonable.

\section{Conclusions}

In this case study, a splat and antisplat detection and visualization method for the analysis of a turbine cascade has been successfully applied.
This was accomplished by adapting the splat detection from 
\cite{Nsonga:2019:splats} to curved surfaces in a curvilinear grid and employing a domain transformation.
It was shown that the resulting visualization, consisting of a surface and a streamline visualization scheme allows an intuitive evaluation of near-wall flows in a turbine.
Based on the results of the presented approach, a qualitative assessment of this flow was performed in the context of the stated research questions.
The parameter study conducted in this work highlights the straightforwardness of parameter selection.
This makes the present approach also viable for other datasets with less researched flow configurations. 

We conclude that the visualization of splats and antisplats is a fitting extension of the toolset required for a profound analysis of turbine flows. It might be used to estimate the amount of heat transfer without actually calculating the temperature transport
and to make preliminary statements on the need for cooling of blade and sidewall.
Another application where splat detection will constitute a useful tool is related to the question of thermal fatigue, e.g. \cite{Dunn_2001, Kuczaj:2010:thermal_fatigue}, a situation where the alternation between hot and cold fluid impinging on a surface is responsible for the fatigue of the material due to its expansion and contraction with high frequency.  

In the future, we will extend the splat detection approach to arbitrary smooth surfaces and unstructured grids thus increasing its versatility and applicability to even more complex situations.
Also, we will investigate improvements in the structural visualization scheme, as clutter becomes challenging for highly turbulent flows.

\acknowledgments{
This work was funded by the German Federal Ministry of Education and Research within the project \textit{Competence Center for Scalable Data Services and Solutions} (ScaDS) Dresden/Leipzig (BMBF 01IS14014B).
%JF Prof. Fr\"{o}hlich and Dr. Ventosa-Molina 
JF and JVM acknowledge funding by DFG under FR1593/15-1 within PAK948. 
}

%{\tt JF: Achtung: in Ref 22 muessen die Abkuerzungen LPT, DNS U-RANS gross geschieben werden.}

%\bibliographystyle{abbrv}
\bibliographystyle{abbrv-doi}

\bibliography{template}
\end{document}